# Factory-simulator: A process-oriented environment for evaluating manufacturing decisions against optimisation-based references

Andrzej Wodecki[a], Wojciech Godlewski[b]*, Michał Chojna[c]

[a] Warsaw University of Technology, Business School, Koszykowa 79, 02-008 Warsaw, Poland

[b] Warsaw University of Technology, Doctoral School, Plac Politechniki 1, 00-661 Warsaw, Poland

[c] Polish-Japanese Academy of Information Technology, Koszykowa 86, 02-008 Warsaw, Poland

* Corresponding author: Wojciech Godlewski. Email: wojciech.godlewski4.dokt@pw.edu.pl

## Abstract

Motivated by discussions with an electronics contract manufacturer about decision support, Factory-simulator provides a deterministic manufacturing environment for controlled research on the business performance of large language model-based multi-agent systems (LLM-MAS). Equal delivery performance can arise from decisions with different financial consequences, complicating evaluation through delivery metrics alone. Its process-oriented model connects procurement and production decisions to order fulfilment and factory finances. Seeded worlds and recorded actions support replay and branching from a shared history to examine alternative decisions and their downstream consequences. Linear and mixed-integer programming models generate profit-oriented reference decisions. Observation-limited comparators use disclosed information; full-foresight references use future events for diagnostic comparison. Controlled examples show equal delivery performance with different profit under the same information and distinguish delivery penalties from deferred payment. Verification includes material and financial balances, replay and forks, and a small allocation instance in which the tested optimisation-based controllers attain an analytically established optimum. An external three-role LLM-MAS client completed four illustrative runs comprising 159 periods, with reconciled balances and exact replay of recorded actions. Matched-world rolling comparisons contextualise the financial outcomes, while full-foresight results provide heuristic reference points rather than certified optima. The software enables researchers to investigate how agent organisation and information access affect business outcomes, including whether distributing procurement and production responsibilities changes profit relative to a centralised controller.

Keywords: LLM-based multi-agent systems; business performance; deterministic manufacturing simulation; optimisation-based reference; process management

**Required metadata**

| Nr | Code metadata description | *Metadata* |
|---|---|---|
| C1 | Current code version | 1.0.0 |
| C2 | Permanent link to code/repository used for this code version | https://github.com/mchojna/factory-simulator/tree/ea92c1bfb28228c8929045e1ddcfa9535da2ea27<br>Release 1.0.0; historical source identity and reproduction data are documented in Section S1. |
| C3 | Legal code license | MIT |
| C4 | Code versioning system used | Git |
| C5 | Software code languages, tools and services used | Python; FastAPI; SQLAlchemy; DuckDB; NumPy; SciPy/HiGHS |
| C6 | Compilation requirements, operating environments and dependencies | Python >=3.14,<3.15; uv >=0.12.10,<0.14. Local verification: macOS arm64, Python 3.14.10. Hosted CI: Ubuntu tests; macOS/Windows frozen-world checks. Locked dependencies and scope: Sections S1.1–S1.2. |
| C7 | If available, link to developer documentation/manual | https://github.com/mchojna/factory-simulator/tree/ea92c1bfb28228c8929045e1ddcfa9535da2ea27/docs |
| C8 | Support email for questions | Wojciech Godlewski; wojciech.godlewski4.dokt@pw.edu.pl |

## 1. Motivation and significance

Recent benchmarks evaluate large language model (LLM) agents in inventory management and business operations [1,2]. Electronics manufacturing coordinates purchasing with production and delivery commitments; semiconductor disruptions can propagate through supply networks [3]. Buying ahead may protect deliveries but ties up cash. Assigning purchasing and production to separate agent roles creates interfaces across which decisions remain coupled. Evaluating LLM-based multi-agent systems (LLM-MAS) [4] therefore requires tracing consequences across functions. InvAgent applies LLM agents to inventory ordering [5]; in Jannelli et al.'s experiments, the lowest-bullwhip configuration was not the lowest-cost one [6]. Factory-simulator separates factory dynamics from agent organisation, allowing studies to vary coordination and information access while holding the factory model, realised world, evaluation objective and horizon fixed.

Here, process-oriented means evaluation across connected business functions [7]. Local functional optimisation can suboptimise the whole [8, Chapter 1]. Galbraith [9] distinguishes slack resources, including buffer stock, from added information-processing capacity as responses to uncertainty, each with its own cost. These frameworks inform the design: Factory-simulator records inventory commitments and storage costs; studies can vary disclosed information and agent organisation.

Related environments use different references. OR-Gym compares reinforcement learning with optimisation and heuristics [10,11]. AIM-Bench uses an ex-post replenishment reference and examines information sharing [2]. InventoryBench reports normalised reward for base-stock, LLM and hybrid procedures [12,13]. EconEvals combines model-based performance references with trajectory analysis [14,15]. The schlably framework supports production-scheduling research with deep reinforcement learning, dispatching heuristics and solver comparisons [16]. Table 1 documents precedents for optimisation references and information interventions. Factory-simulator combines these capabilities with receivables-inclusive profit and branching from recorded histories.

*Table 1. Evaluation support in related environments and Factory-simulator. Selection covers production, supply-chain and business-operation tasks. Entries describe documented capabilities; unreported capabilities are not treated as absent. Information parity is protocol-specific. Section S2 gives sources and a wider comparison.*

| Environment | Reference or evaluation target | Information, control and diagnostics |
|---|---|---|
| OR-Gym [10,11] | MILP and heuristics; inventory-task shrinking-horizon LP and perfect-information oracle; task rewards. | Inventory state and action history; task-specific optimisation models. |
| AIM-Bench [2] | Ex-post replenishment reference; ordering deviations and inventory costs. | Multi-agent settings and information-sharing treatments; period-level ordering behaviour. |
| InventoryBench [12,13] | Capped base-stock baseline; normalised reward for OR, LLM and hybrid methods. | Pregenerated demand and lead times; historical and contextual inputs; recommendation or forecast flows. |
| Vending-Bench [1] | Long-running business operation; final net worth and a single-run human baseline. | Agent trajectories, tool use and simulated supplier correspondence. |

| Environment | Reference or evaluation target | Information, control and diagnostics |
|---|---|---|
| schlably [16] | DRL agents compared with dispatching heuristics and a solver; reward and gap to the solver solution. | Generated scheduling instances, Gym environments, logged outcomes and Gantt charts. |
| SCML [17,18] | Comparisons of negotiating agents in supply-chain worlds. | Configurable contexts and controlled world conditions for behavioural comparisons. |
| MLPro-MPPS [19,20] | User-configured production systems for learning and control. | Sensor and component states; integration with reinforcement learning and game theory. |
| prodsys [21] | Production configurations; throughput, throughput time, work in progress and reconfiguration cost. | Seeded simulations and stored events for post-processing. |
| SimPROCESD [22] | Maintenance and production policies; net production profit and quality. | Equipment condition and resource-use indicators. |
| Factory-simulator (this work) | LP/MILP-based comparators; full-foresight diagnostic reference; horizon profit including receivables and stock salvage. | Recorded worlds and history forks; configurable disclosure, with access enforced by the harness; shared material and financial accounting. |

SupChain-Bench evaluates knowledge and tool-mediated retrieval against ground truth [23,24].

A common engine applies the same rules to submitted policies. Researchers can trace costs behind equal service and separate profit effects from payment timing (Section 3).

## 2. Software description

### 2.1 Software architecture

The engine owns state transitions and accounting (Fig. 1). An HTTP service exposes a JSON API and persists worlds and run records. A benchmark layer supplies LP/MILP-based comparators, diagnostic references and economic scores. The included single-prompt, tool-guided and coordinated clients can be extended with study-specific roles. The engine exposes one controller interface per run. General agent frameworks such as mango provide roles, messaging and coupling to external simulators [25]; Factory-simulator leaves agent organisation to the external harness.

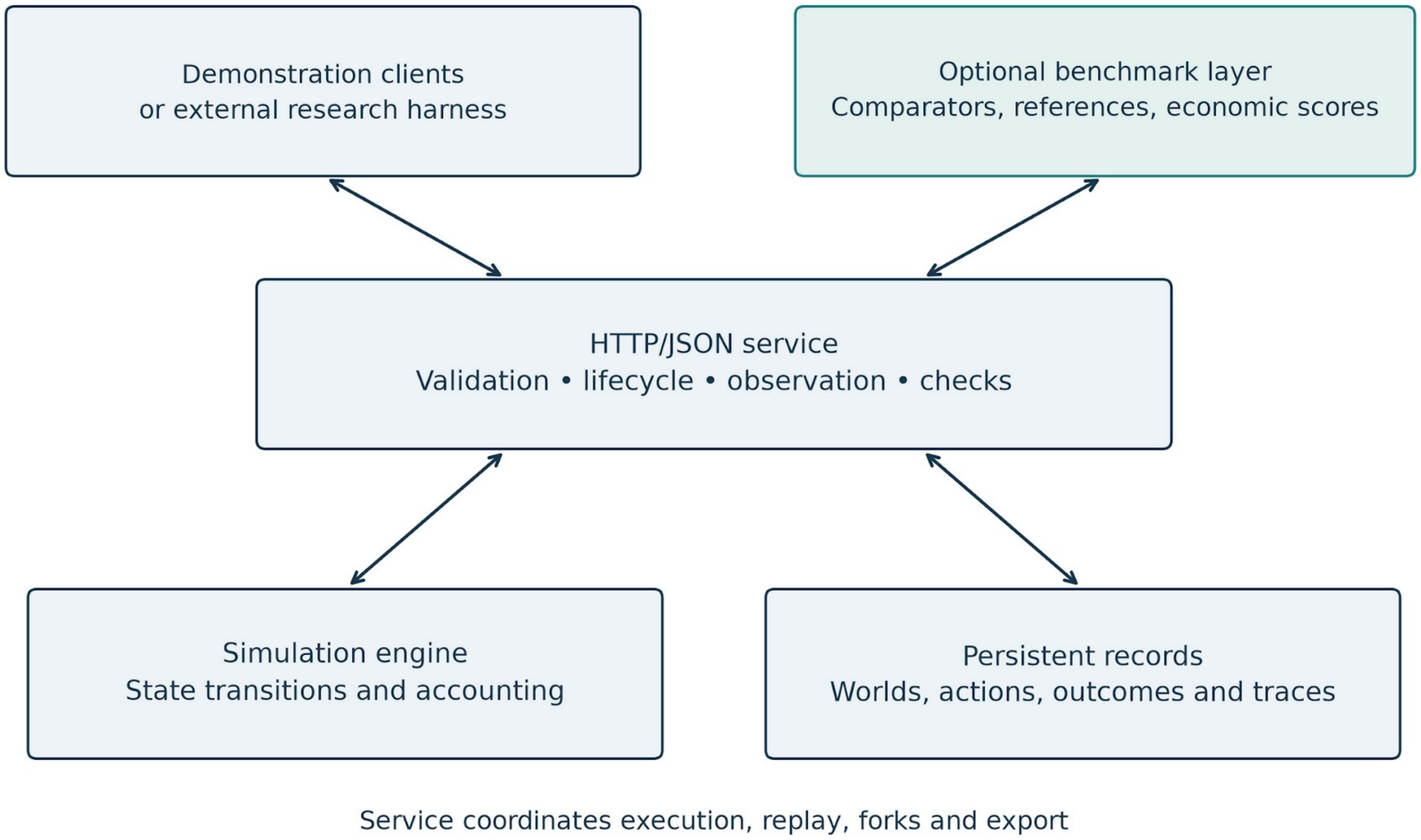

*Fig. 1. Architecture supporting decision evaluation. Demonstration clients accompany the engine; an external research harness defines the study and controls access. Codex (OpenAI, GPT-6 Astra) assisted with the Python/Matplotlib drawing code.*

A manifest defines factory entities, recipes and parameter distributions within the simulator's modelling assumptions; a scenario specifies initial conditions, the horizon and disturbances. A seed determines the realised world. The electronics factory included in the release (the library factory) is a stylised example with synthetic parameters, not calibrated to company data. It contains three products, six components and nine materials, with four suppliers and four customers. Four production lines and two warehouses connected by a transfer route impose operational constraints. Alternative recipes and purchasing discounts introduce choices beyond production volume.

A client registers a manifest and scenario, starts a run, and retrieves its description and observation. API observations report cash, receivables, stock, open customer and supplier orders and line states, with current prices, disruptions and forecasts where available and disclosed. The client submits decisions, optionally after feasibility checks; the service advances the period. Records and history forks remain accessible through the same interface. Section S1.1 includes an API-client excerpt.

### 2.2 Process model and execution

Material dependencies connect purchasing with delivery commitments (Fig. 2). Warehouses constrain storage, equipment condition affects production, and payment terms connect physical execution to liquidity. Customers receive goods only through explicit shipment decisions, making order selection part of the controller's responsibility. Maintenance and warehouse transfers compete with other uses of resources. This structure permits a locally attractive action to have adverse consequences elsewhere in the process.

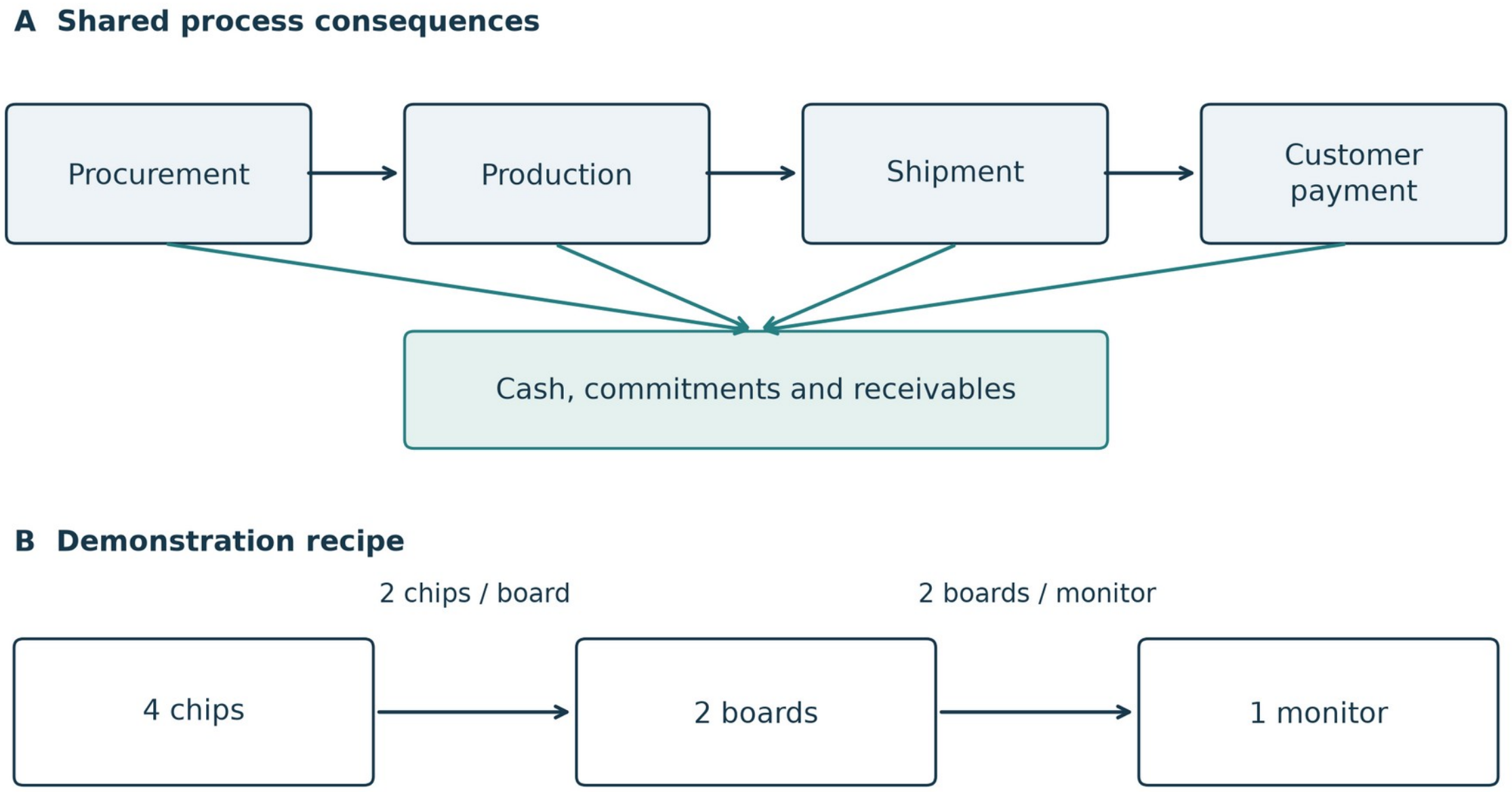

*Fig. 2. Process relationships and a simplified bill of materials. The two-stage chain is used in Sections 3.1–3.3 and integration case A3; the other integration cases use the larger library factory. Shipment and customer payment are separate events. Codex (OpenAI, GPT-6 Astra) assisted with the Python/Matplotlib drawing code.*

Time advances in discrete periods (Fig. 3). Deliveries and events precede observations; decisions then execute before the period closes. Production with lead time one enters stock at closing and can be used in the next period. Shipments execute first, followed by cancellations, transfers and repairs. Due production plans precede current production, and purchases execute last. Future production plans do not reserve resources. Each action is accepted or rejected as a whole against the updated state, so competing actions require an explicit order.

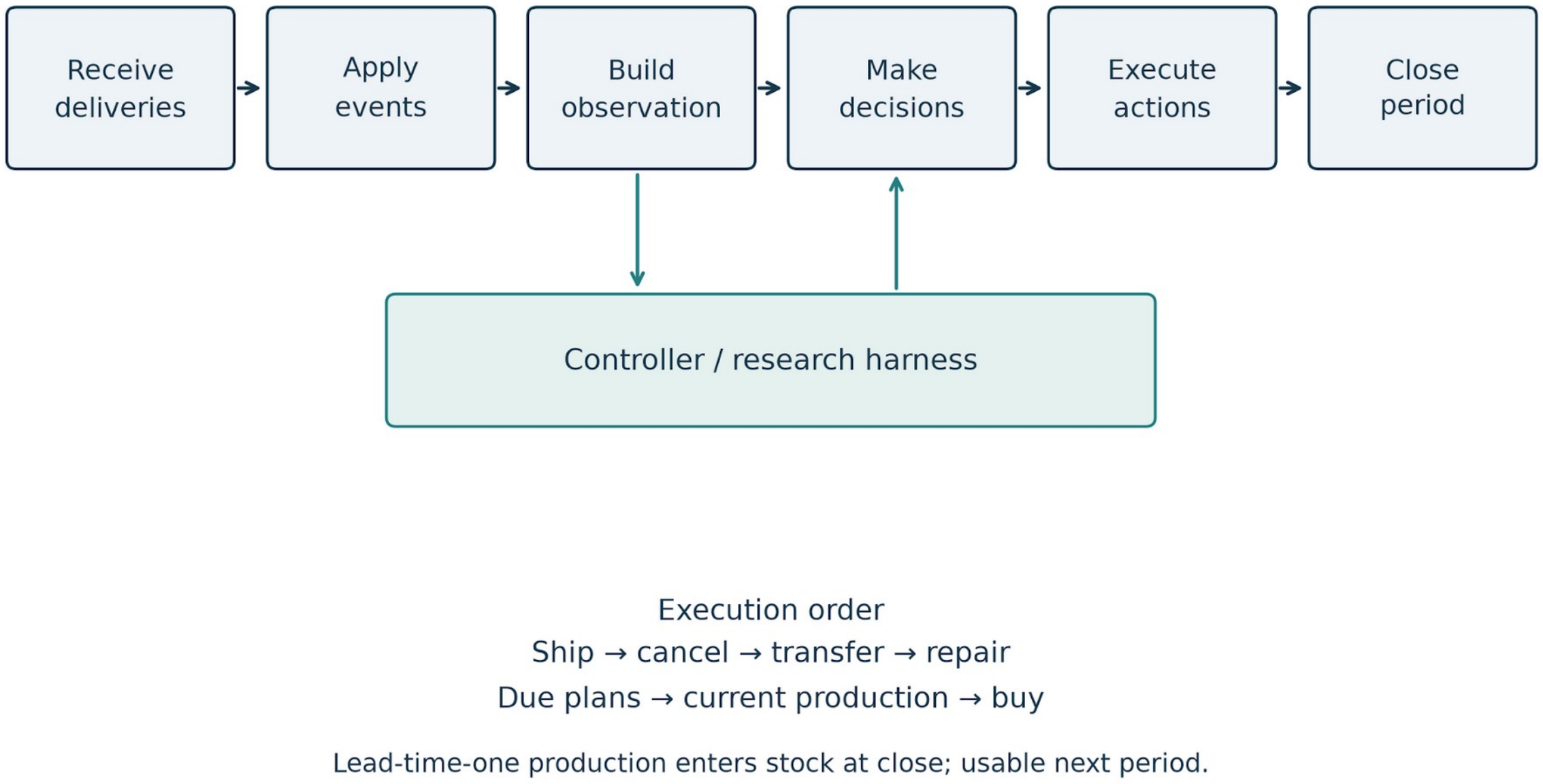

*Fig. 3. One decision period. The harness receives a description and current observation, then returns actions. Within-period execution order and production timing are model assumptions. Codex (OpenAI, GPT-6 Astra) assisted with the Python/Matplotlib drawing code.*

The credit check admits a purchase only when its cost does not exceed cash plus the credit limit less existing purchase commitments. Other costs can take cash below that bound; the library factory charges interest on debt and excess borrowing. Monetary calculations use decimal arithmetic. These conventions and the fixed action ordering are part of the decision problem and must be respected when translating an aggregate plan into actions.

### 2.3 Scenarios and information

The library contains a steady-market baseline and 22 scenarios spanning six families (Table 2). The horizon is configurable; supplied scenarios run for 52 periods. They connect operational constraints to business questions about procurement and fulfilment. Section S3 describes every scenario, its rationale and suitable diagnostics. These synthetic business conditions support controlled studies of a stylised factory. Industrial calibration and calibration of scenario difficulty remain outside the present scope.

*Table 2. Scenario families beyond the steady-market baseline. Questions indicate possible studies, not established performance results.*

| **Family (count)** | **Illustrative conditions** | **Research question** |
|---|---|---|
| Demand (6) | Seasonality, growth, decline, surge and product life cycle | When should stock be built or reduced? |
| Suppliers (5) | Outages, delays, quality loss, shortage and price changes | How should procurement adapt? |
| Production lines (4) | Breakdowns, capacity cuts and yield loss | How do recovery decisions affect fulfilment? |

| Family (count) | Illustrative conditions | Research question |
|---|---|---|
| Cash and space (2) | Credit pressure and a full warehouse | Which action releases the binding resource? |
| Information (3) | Forecasts, advance notice and hidden parameters | How does disclosure alter decisions? |
| Combined disturbances (2) | Regional crisis; demand surge with breakdown | How do interacting disruptions change priorities? |

Seasonality motivates stock-building subject to storage limits; advance price information can favour purchases that consume credit. In the full-warehouse case, intermediate production can release space for missing inputs. Combined demand and capacity shocks create conflicting priorities.

Descriptions and observations are derived from the underlying world. Scenarios can hide parameter groups, supply noisy forecasts or disclose a disruption in advance. Some information scenarios also change demand; the documented protocol specifies matched controls. The world endpoint requires the requested run to be finished, but the service does not enforce controller-specific access restrictions. The harness must restrict access to scenario configurations and to completed runs that share the world, such as a fork's parent, and provide comparable access to feasibility checks. Equal information means a common disclosure regime and permitted tools; observations can diverge after different actions.

### 2.4 Recorded worlds and alternative decisions

For fixed worlds and actions, execution is deterministic within a compatible software environment. Exogenous draws use separate named streams generated before execution. Decisions alter consequences, not the shared draws. Records preserve actions, outcomes and execution traces. Replay first checks package, Python and NumPy version compatibility, then compares regenerated execution with the records.

A completed run can be forked at a chosen period. The service replays the preceding closed history and accepts an alternative continuation on the same exogenous world (Section 3.2). The comparison includes downstream policy responses; the harness must manage controller memory and continuation rules.

### 2.5 Evaluating decisions against reference outcomes

An evaluation specifies its objective and horizon before comparing controllers. The supplied optimisers pursue profit through planning-window models and terminal-value approximations; they are not service-optimal references. A realised result measures performance on one world; assessing a policy under uncertainty requires a defined population of worlds. Horizon profit is given by Equation (1).

$$\pi = C_h - C_0 + R_h + V_h. \qquad (1)$$

Here, $C_0$ and $C_h$ are initial and final cash, $R_h$ is outstanding customer receivables at nominal value, and $V_h$ is warehouse-stock salvage value. Work in progress and goods in transit have no terminal salvage value. Initial stock incurs no acquisition charge. Undelivered supplier orders and customer orders not yet due receive no terminal valuation (Section S4.2). Profit should be reported alongside service measures with explicit denominators and timing, because a delivery rate and an economic outcome answer different questions.

A score normalised by the no-decision baseline and full-foresight reference is available as a diagnostic, not an optimality percentage (Section S4.2). Comparisons under common disclosure should report direct profit and service differences against an observation-limited comparator on matched worlds.

The full-foresight reference plans with the realised future world, repeatedly solving an LP or MILP and adapting decisions to engine constraints. Later-period binary choices can be relaxed, and execution includes heuristics. Its realised profit has no general guarantee of global optimality. Solver statuses and gaps concern the planning program. A shortfall against this reference combines differences in information and planning procedures.

Observation-limited comparators use disclosed descriptions and observations, with feasibility-check access (Section 2.3). They include full-horizon limited, rolling, buffered rolling, forecast-aligned and myopic controllers (Section S4). An informed diagnostic additionally receives hidden parameter means, without future draws (Fig. 4). Matched interventions separate information effects from algorithmic choices. An exact-integer control removes one planning relaxation for small programs; an engine-level optimum requires independent proof, as in Section 3.3.

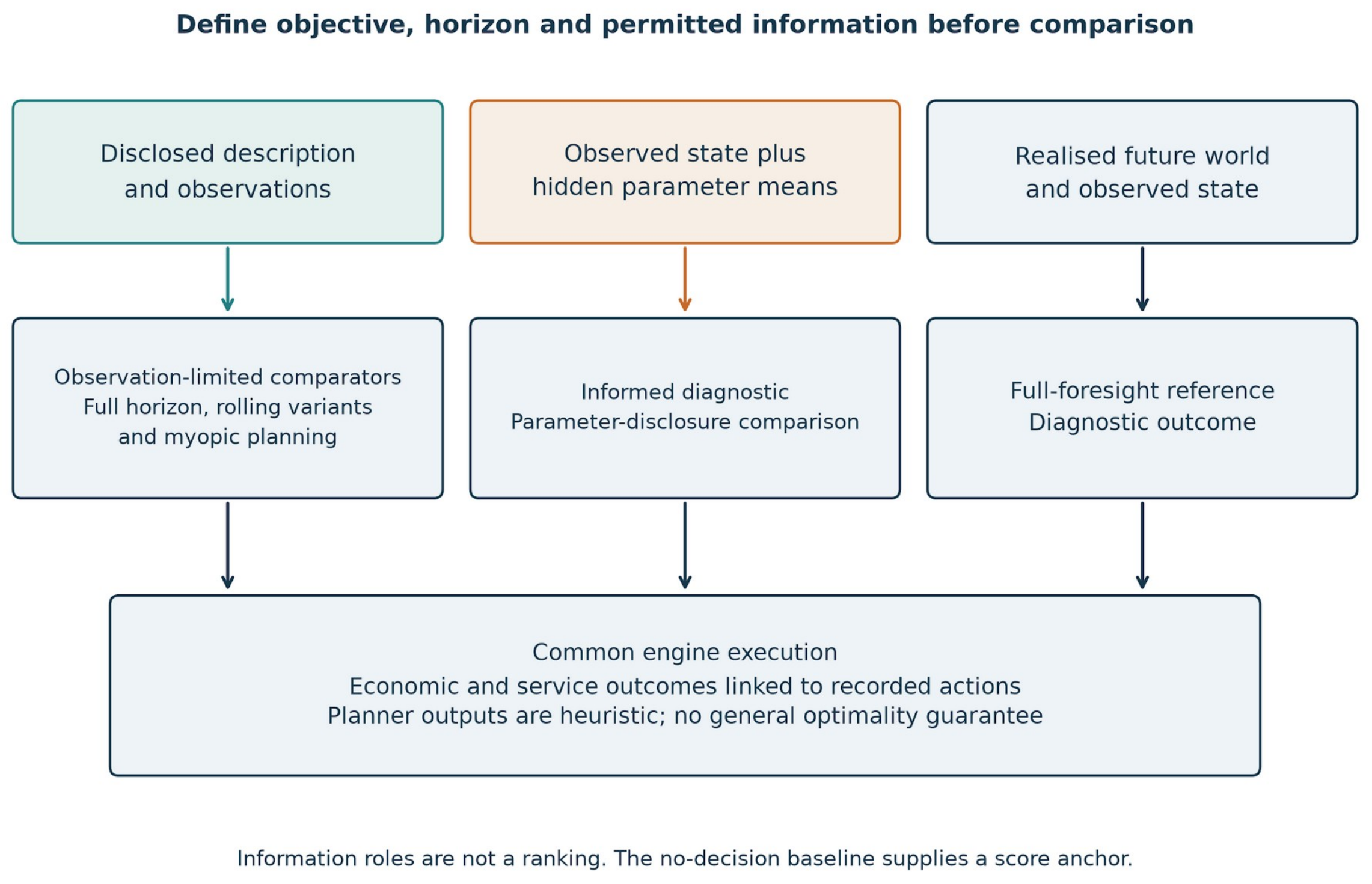

*Fig. 4. Roles of comparison controllers and diagnostic references. Available information defines each role; the columns do not imply a performance ordering. Engine execution determines all outcomes. Feasibility-check access must also be specified. Codex (OpenAI, GPT-6 Astra) assisted with the Python/Matplotlib drawing code.*

## 3. Illustrative examples

Sections 3.1–3.3 use small deterministic factories with controlled decision policies to demonstrate the evaluation logic. Section 3.4 summarises verification; Section 3.5 applies these evaluations to a live LLM-MAS. Supplement S1 supplies inputs and reproduction instructions; section numbers prefixed S refer to it.

### 3.1 Equal service, different economic outcomes

A ten-period electronics-assembly factory serves demand for four monitors per period. Each monitor requires two boards and each board two chips (Fig. 2). Unit costs are 0.5 for chips, 1 for board production, 2 for monitor assembly and 1 for shipment; monitors sell for 50. Yield is one, and supplier and production lead times are one period. Customers pay immediately; lost sales cost 10 per unit. Initial cash is 1000, with 16 chips, eight boards and four monitors. Storage costs per unit and period are 0.0023, 0.0090 and 0.0270 for chips, boards and monitors; salvage values are 0.1, 0.5 and 2. Interest and wear are zero. Chip purchases are blocked in periods 3–6, disclosed either at onset (period 3) or by advance notice in period 1.

Without advance notice, rolling, buffered rolling and full-horizon limited controllers each deliver 24 of 40 monitors on time and earn 919.17 (Fig. 5). With advance notice, rolling delivers 28 and earns 1130.96; buffered rolling and full-horizon limited each deliver 40, earning 1732.43 and 1766.08. At these storage rates, advance notice increases realised profit by about 212 for rolling and 847 for full-horizon limited; its effect depends on the controller in this instance. Buffered rolling and full-horizon limited achieve equal service under the same information but differ in profit by 33.66 (from unrounded results). Additional material purchases and storage charges, partly offset by salvage, account for the difference. Their policies differ in both planning horizon and terminal stock valuation. The full-foresight reference delivers 40 and earns 1766.08 in both regimes, matching full-horizon limited with advance notice in this instance.

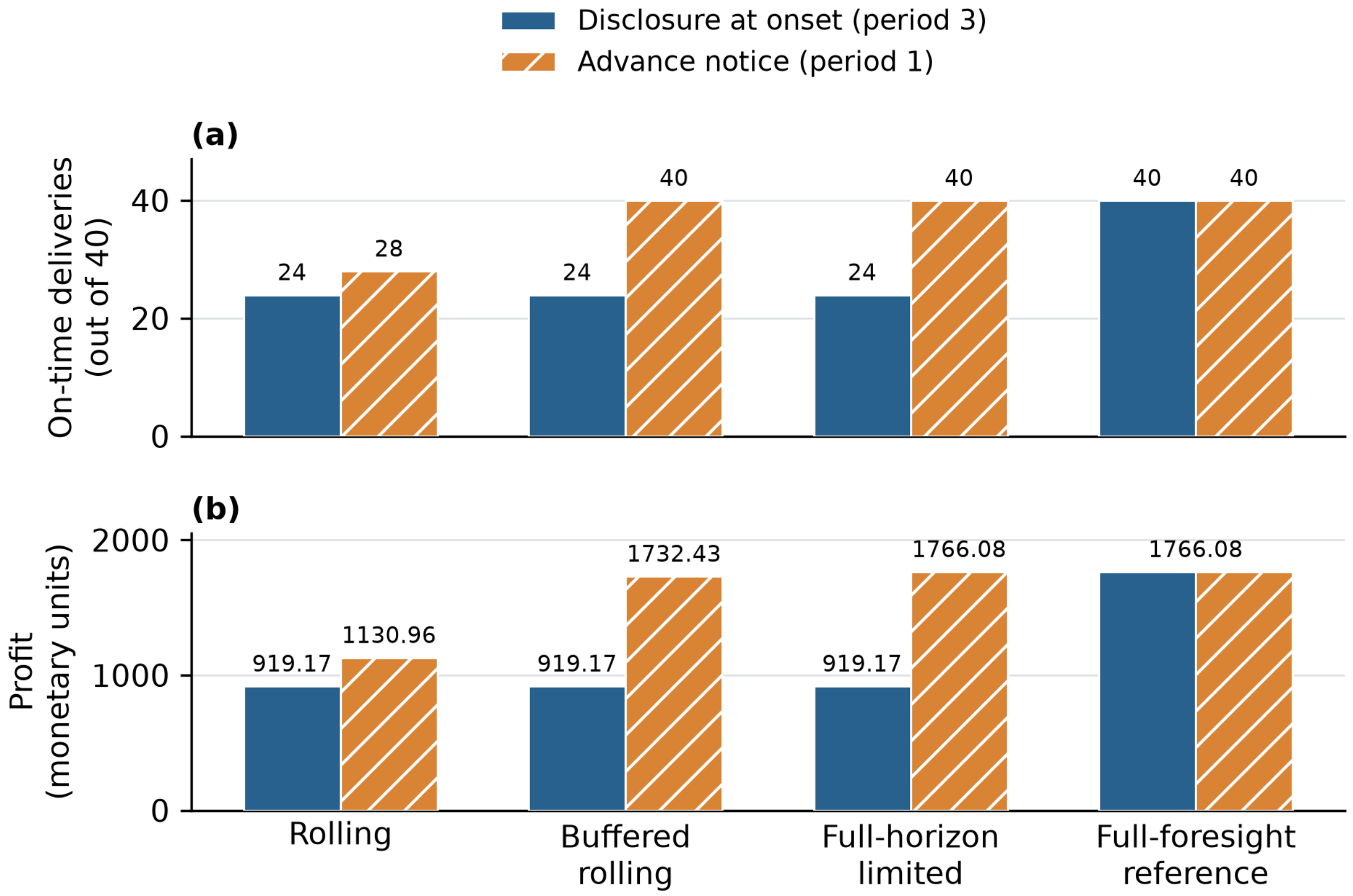

*Fig. 5. Timely deliveries and profit under disclosure at onset (period 3) and advance notice (period 1), with positive storage costs (Section 3.1). With advance notice, buffered rolling and full-horizon limited deliver all 40 monitors but differ in profit by 33.66 (from unrounded results). Monetary units denote model monetary units. Controller definitions are in Section S4.*

With advance notice, rolling's deliveries depend on the storage rate: 28 at the rates above (about 0.45% of replacement cost per period) and at the tested 1%, 2% and 5%, but 40 at 0.02%, 0.1% and 0.2%. Section S5.2 reports the controlled continuation, the sweep, the zero-cost variant and reconciled outcomes.

### 3.2 Tracing an alternative decision through the process

A five-period factory uses the same recipe and unit costs, starting with cash 1000 and no stock. A customer orders four monitors due in period 4, tolerates one period of delay and pays one period after shipment. Late delivery costs 2 per unit per period; lost sales cost 10 per unit. Yields are one and lead times one period; storage charges, interest and wear are zero. The controller buys 16 chips in period 1, makes eight boards in period 2 and assembles four monitors in period 3.

A standing rule ships available goods against open orders. Shipment occurs in period 4 and payment in period 5. Revenue of 200 less total costs of 28 gives profit 172 and final cash 1172 (Fig. 6). A fork beginning in period 3 postpones assembly to period 4 while retaining the shipping rule. Shipment moves to period 5, incurs a late penalty of 8, and leaves payment beyond the horizon.

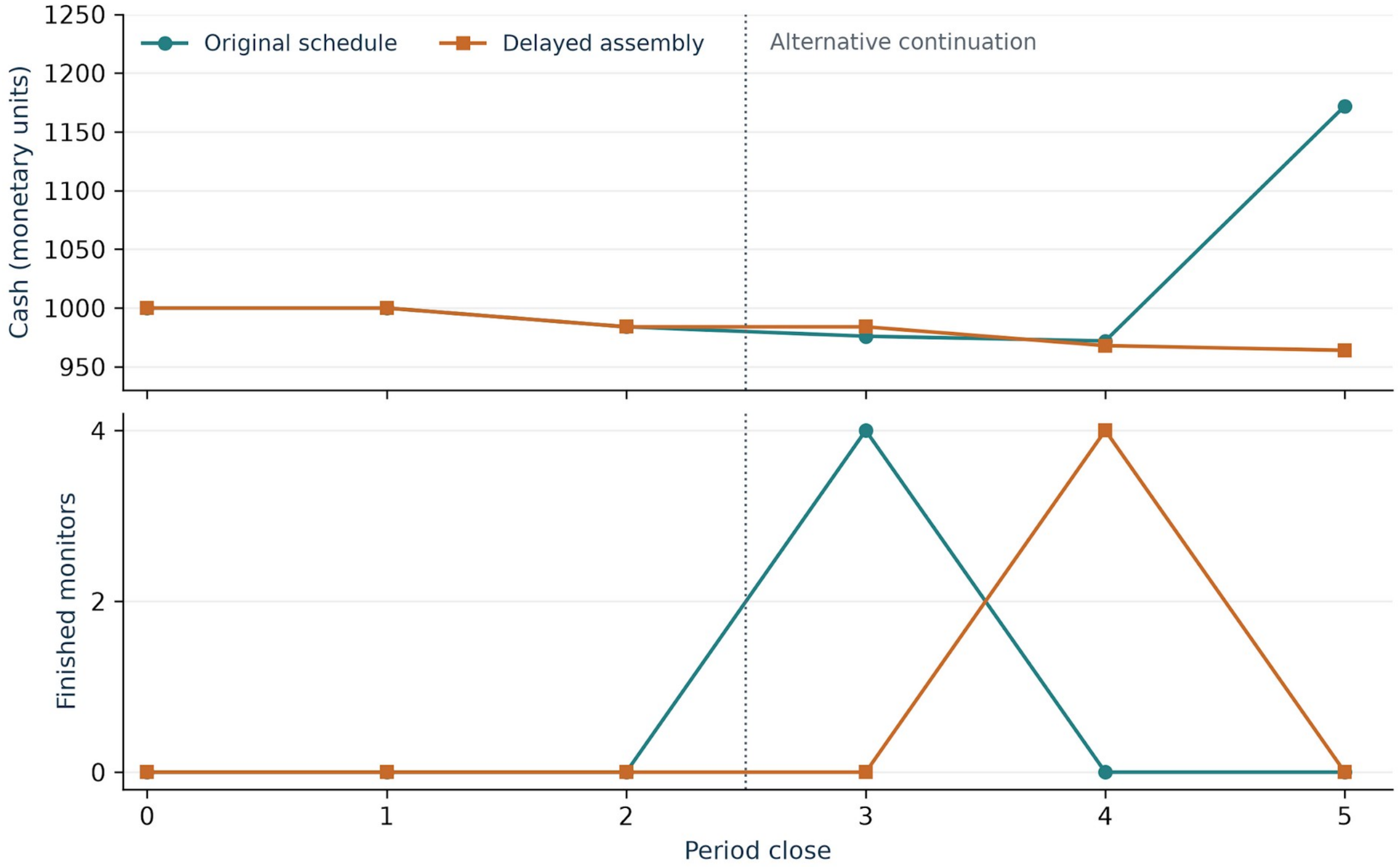

*Fig. 6. An alternative assembly decision propagates to shipment and payment. The common closed history ends in period 2. Final cash differs by 208, while profit differs by 8. Monetary units denote model monetary units. Period 0 is the initial state; lines connect discrete observations.*

Final cash in the fork is 964 and receivables are 200, giving profit 164. The cash difference of 208 comprises delayed payment of 200 and an economic penalty of 8. With zero interest and nominal receivables, this example separates the delivery penalty from payment timing. The comparison includes the downstream response of the unchanged shipping rule. A no-decision run loses 48: four units incur a late penalty of 8 and a terminal lost-sale penalty of 40 (Section S5.1).

### 3.3 A checkable optimal allocation

A one-period micro-instance starts with 0.001 monitors and two orders of that quantity, priced at 50 and 70 per unit. Shipping costs 1 per unit; unserved demand incurs a penalty of 10, and salvage is zero. At the implemented precision, the three possible allocations are no shipment, the lower-price order and the higher-price order, yielding −0.020, 0.039 and 0.059. Production completes after shipment and cannot improve same-period revenue. The upper bound is $0.001 \times (70 - 1) - 0.001 \times 10 = 0.059$. The full-foresight reference, exact-integer control and rolling comparator attain it in this instance. Section S5.3 gives the proof and precision checks.

### 3.4 Verification scope

Hosted CI passed Ubuntu tests and frozen-world checks on macOS and Windows (Section S1.2). Across staged macOS runs, all 3430 distinct non-stress tests and doctests passed cumulatively. Tests cover engine behaviour, API handling, replay, forks and simulated-model integration. Material and financial reconciliation passed for seven fulfilment and 47 notice-analysis runs; seven allocation runs passed financial reconciliation.

Following Sargent [26], these results provide targeted software-verification evidence. Developmental scenario calibration documented with the software is separate from this verification and from industrial validation.

The release documentation reports six scenarios failing outcome-direction checks for specified controller–seed pairs. Such failures alone do not diagnose a broken scenario mechanism or establish task difficulty; Section S3 distinguishes these questions.

### 3.5 Integration with an LLM-based multi-agent controller

An external extension of the repository's coordinated client connected planner, procurement and reviewer roles to the HTTP/JSON service, using OpenAI gpt-6.1-terra with low reasoning effort on 12–13 November 2026. It allowed up to two additional reviewer revisions per period and included a filter that removes rejected decisions. Table 3 reports all four final runs. Supplement S1 documents the final protocol and retains earlier attempts in Section S7.

All 159 periods completed, with 483 model responses including six additional revisions. The filter and transport retries were unused; C1 and C2 each required three additional reviewer revisions. Period-level material and financial checks found no discrepancies, and replay reproduced recorded actions and outcomes without further model calls. Replay does not reproduce stochastic decision generation.

C1 and C2 shared a world and disclosure settings, with fresh memory. C1 ended with positive profit, negative cash and outstanding receivables of 211964.41. C2 paid 3783.17 in interest and ended below its credit limit, which constrains purchase admission rather than every debit. These records distinguish profitability from liquidity.

Matched-world outcomes provide context (Table 4). The outage-with-notice world adds a chip-supply disruption in periods 10–41, disclosed from period 6, to the steady-market world; this comparison does not isolate the value of notice. Rolling uses disclosed observations and feasibility checks; full foresight also uses future events. Two full-foresight solves in that world reached the 30-second limit, retaining usable MILP solutions (Section S5.5). B3 shipped 21.71% of ordered units on time against 65.85% for rolling. In the outage-with-notice world, C1 shipped 54.7% of ordered units against 45.1% for rolling, but 18.1% on time against 36.7%, and earned more; C2 earned less than rolling. The realised profits of C1 and C2 differed by about 211571 model monetary units. These runs do not establish a controller ranking.

*Table 3. Final LLM-MAS runs: three periods for A3, 52 otherwise. Financial values and differences from matched-world rolling are in model monetary units. A3 illustrates small-factory integration; no rolling comparison is reported.*

| Run | Scenario | On-time / ordered (%) | Horizon profit | Final cash | Δ profit vs rolling |
|---|---|---|---|---|---|
| A3 | Small factory | 66.67 | 328.00 | 1328.00 | — |
| B3 | Steady market | 21.71 | 552071.45 | 361309.71 | −316898.86 |
| C1 | Outage with notice | 18.11 | 203626.76 | −15502.68 | +24167.97 |
| C2 | Outage with notice | 15.12 | −7944.52 | −119243.39 | −187403.32 |

*Table 4. Matched-world horizon profits, seed 7. C1 and C2 share the outage-with-notice world. No-decision and myopic actions and cash flows coincide across worlds (Supplement S1). Myopic plans only for the current period, with terminal stock and pipeline valuation (Section S4). The heuristic full-foresight reference uses additional information (Section 2.5).*

| World | No decisions | Myopic | Rolling | Full foresight |
|---|---|---|---|---|
| Steady market | −709331.23 | −432814.34 | 868970.31 | 1098988.81 |
| Outage with notice | −709331.23 | −432814.34 | 179458.79 | 892273.68 |

## 4. Impact

Factory-simulator lets researchers examine what a given level of service costs in the modelled factory and where those costs arise. In Section 3.1, two policies with the same information deliver all 40 monitors on time but differ in profit through material purchases and storage charges, partly offset by salvage. Section 3.2 traces an assembly delay through shipment and collection. The live integration adds auditable LLM-MAS trajectories and matched-world comparisons with rolling control.

Future studies can ask: How does dividing procurement and production between agent roles change profit relative to centralised control on the same world? How do inventory policies and information sharing affect factory profit and separately measured inference costs (Section S4.1)? How does changing one decision affect outcomes under a specified continuation policy, using a fork from a common prefix? Factory-simulator is the planned factory environment for doctoral research at Warsaw University of Technology on decision-making supported by LLM-MAS. The API also permits heuristic or human-controlled policies.

The library factory is stylised, and industrial calibration and independent adoption have not been demonstrated. Discrete periods and fixed execution order limit detailed scheduling applications; the financial conventions and evidence boundaries are specified in Section 2.5 and Section S6.

## 5. Conclusions

Factory-simulator links manufacturing decisions to business outcomes through optimisation-based references and replayable histories. Controlled examples distinguish service, profit and payment timing. With an external client, a live LLM-MAS demonstration completed four runs with reconciled balances and exact recorded-action replay. Matched-world references contextualise these outcomes. Comparative effectiveness requires repeated studies, for which Factory-simulator provides a common environment.

## Code and data availability

Factory-simulator 1.0.0 is available under MIT [27], including the external client and reproduction scripts. The data and reports are publicly available in GitHub Releases under CC BY 4.0, with the fixture exception specified in the archive. Section S1 gives versioned links and checksums.

## CRediT authorship contribution statement

Andrzej Wodecki: Conceptualization, Supervision, Project administration. Wojciech Godlewski: Writing – original draft, Visualization. Michał Chojna: Software (design, architecture and testing). All authors: Methodology, Software, Validation, Formal analysis, Investigation, Resources, Data curation, Writing – review & editing.

## Funding

This research did not receive any specific grant from funding agencies in the public, commercial, or not-for-profit sectors.

## Declaration of competing interests

The authors declare that they have no competing interests.

## Declaration of generative AI and AI-assisted technologies in the manuscript preparation process

The authors used Claude (Anthropic) and Codex (OpenAI) for drafting, English-language editing, reference checks and reproducibility checks. Codex assisted with code development and review and with preparing figure-generation scripts. The authors reviewed and edited the outputs and take full responsibility for the article.

# Supplement S1

Factory-simulator: A process-oriented environment for evaluating manufacturing decisions against optimisation-based references

Andrzej Wodecki, Wojciech Godlewski, Michał Chojna

This supplement documents Factory-simulator 1.0.0 and its demonstration evidence. Code is maintained in the repository; Factory-Simulator-Reproduction.zip contains data, reports and figures only. The archive is publicly available at https://github.com/mchojna/factory-simulator/releases/tag/factory-simulator-reproduction. README.txt gives commands; CONTENTS.tsv, MIGRATION.tsv and PROVENANCE.json identify retained records and their original locations.

## S1. Software metadata and reproduction

| Metadata item | Value |
|---|---|
| Software and licence | Factory-simulator 1.0.0; MIT |
| Source repository | https://github.com/mchojna/factory-simulator |
| Published code snapshot | ea92c1bfb28228c8929045e1ddcfa9535da2ea27 |
| Software release and data archive | Code: https://github.com/mchojna/factory-simulator/releases/tag/v1.0.0 (commit ea92c1bfb28228c8929045e1ddcfa9535da2ea27). Data: https://github.com/mchojna/factory-simulator/releases/tag/factory-simulator-reproduction. Archive SHA-256: 3340137bc74bb754ded7b350db4fcf6c0b5e7b782b2113d5fb09abadb5d67d8f. |
| Language and environment | Required Python >=3.14,<3.15. Verified environment: Python 3.14.10; macOS 26.5.2 arm64; uv 0.13.6 for the original reproduction and 0.13.7 for supplementary checks. |
| Numerical software | NumPy 2.5.3; SciPy 1.18.1; bundled HiGHS 1.12.0 |
| Service and persistence | FastAPI 0.142.2; SQLAlchemy 2.1.3; DuckDB 1.5.6 |
| Documentation | docs/README.md in the repository |
| Execution/test dependencies | Repository uv.lock; default development dependencies. Historical installed versions are in historical/v06-verification/environment.json. |
| Figures | Matplotlib 3.11.2; editable SVG and vector PDF plus PNG |
| Support contact | Wojciech Godlewski; wojciech.godlewski4.dokt@pw.edu.pl |

The published tag v1.0.0 identifies commit ea92c1bfb28228c8929045e1ddcfa9535da2ea27. Earlier development snapshots used the same version label; use the exact commit in Section S1.1. Original scientific records used engine commit 319489ac6f7c45ae022521681b1a93b5e9e01ecd; validation/repository-2026-11-19/ reports describe commit 394f68c6f16da1a280cfb0cef3a70d3fa46e066a. These commits belong to the earlier development history and are not public checkout targets. All 185 files in src/ and uv.lock match the historical source hashes in historical/source/source-distribution.json. REPRODUCTION.json distinguishes the published tag from historical validation; SHA256SUMS.json covers the data payload. Replay requires compatible package, Python and NumPy versions. Controlled examples use repository test fixtures independently of model clients. Their solve limit is 30 seconds; none reached it. Section S5.5 separately reports library-world solver statuses. Timings are excluded from deterministic record comparisons; full-archive byte identity across executions is not claimed.

### S1.1 Installation and rerunning the demonstrations

Clone the repository and check out the exact reproduction commit before installing Python 3.14.10 with uv >=0.12.10,<0.14. Download Factory-Simulator-Reproduction.zip beside the checkout. The commands below use a POSIX shell, default locked development dependencies and new output directories. No API keys are required. Keep generated outputs outside the extracted data archive; never run with Python assertions disabled.

```
git clone https://github.com/mchojna/factory-simulator.git

cd factory-simulator

git checkout ea92c1bfb28228c8929045e1ddcfa9535da2ea27

uv sync --locked --python 3.14.10

uv run python -m scripts.reproduction.prepare_reproduction --archive ../Factory-Simulator-
Reproduction.zip --output data/reproduction

PACKAGE=data/reproduction/Factory-Simulator-Reproduction

uv run python -m scripts.reproduction.check_recorded_inputs --package-root "$PACKAGE"

uv run python -m scripts.reproduction.replay_saved_actions --case all --package-root "$PACKAGE"
--output data/replay-check

uv run python -m examples.coordinated_planning.run --offline --case all --archive ../Factory-
Simulator-Reproduction.zip --output data/client-check

uv run python -m scripts.reproduction.article all --data "$PACKAGE" --output data/article-check
```

The article entry point runs the notice matrix, sensitivity and continuation checks, order-fulfilment/fork examples and micro-allocation examples, then renders all figures. It verifies the archive before and after execution and blocks external network calls. The examples use the repository's in-process ASGI test client. On 19 November 2026, 34 notice cases, 12 sensitivity cases, one controlled continuation and 35 small-example record comparisons passed; all six figures were generated. Results, period records and worlds agreed after JSON parsing. Solver diagnostics use the documented tolerance, excluding timings. Reports are in validation/repository-2026-11-19/.

The following abridged extracts illustrate registration and run creation from scripts/reproduction/article/run_allocation.py in the repository. The full script supplies the benchmark client c, manifest m, scenario s, imports and controller. They are not a standalone client; use the complete entry point above.

```
def setup(c,m,s):

 mid=b.send(c,'POST','/manifests',m)['id']

 return b.send(c,'POST',f'/manifests/{mid}/scenarios',s)['id']

def play(c,sid,agent_kind,base=None,policy=None):

 r=b.start_run(c,sid,7); check=partial(b.create_check,c,r['id'])
```

The omitted remainder of play selects the controller, submits its decisions and reads the recorded outcomes. These extracts are from an executed in-process ASGI client; they do not constitute a separate network-transport test.

### S1.2 Verification scope and reproduction

The published code passed all 13 jobs in the linked CI run for the commit identified in Section S1.1. Ubuntu checks include ordinary tests and eight property/slow shards; macOS and Windows run the frozen-world selection. CI also checks package building, clean installation and pre-commit rules; it does not run the complete test suite on every platform.

Archived local reports document 3430 distinct non-stress test items passed across staged macOS runs for the historical engine commit (Section S1). This cumulative coverage is separate from CI; the stress campaign was not repeated for this article. Test selections, commands and reports are retained in historical/v06-verification/ and historical/verification/ in the data archive; Section S7.1 describes the execution conditions.

End-to-end tests cover no-decision and optimisation clients, feasibility checks, resume, fork and record inspection over loopback HTTP. Model-adapter tests use simulated transport; live LLM-MAS demonstrations and their checks are reported separately in Section S5.4.

To reproduce the article results, obtain the code and data archive and follow the commands in Section S1.1 and scripts/reproduction/README.md in the repository. These cover archive verification, recorded-input checks, saved-action replay, offline client checks, re-execution of controlled examples and figure generation, without model calls. Sections S5.4–S5.5 distinguish replay from new model sampling and optimisation.

### S1.3 Figures and the notice example

```
uv run python -m scripts.reproduction.article figures --data "$PACKAGE" --output data/figure-check
```

Run the command from the repository root. The data archive contains six current figures in PNG, PDF and SVG, without duplicated rendering assets. Plotting code is in scripts/reproduction/article/. Figures 1–4 are explanatory diagrams drawn by Python/Matplotlib; Figure 6 uses archived example records and Figure 5 uses notice/recorded-results.json. Codex (OpenAI) assisted with the plotting code for Figures 1–6; the images were rendered by code, not directly generated by a model. GPT-6 Astra was used through Codex for the drawing code of Figures 1–4. These disclosures are unchanged by the code/data migration.

The notice data for Section 3.1 and Figure 5 are in notice/ of the data archive identified in Section S1; the executable procedures are in scripts/reproduction/article/ in the repository.

The notice command group reruns the 34-case matrix, twelve additional sensitivity cases, a controlled continuation and planning diagnostics. Use the all command in Section S1.1, or substitute notice to run only this group. Historical checks on the original macOS workstation and Linux container are retained with their original scope. The 19 November macOS repository validation again matched result, period and world fields; six reconstructed planning inputs and 42 diagnostic solves passed the documented numerical checks. No model credentials are required. Section S5.2 gives numerical evidence and Section S5.5 describes the historical Linux check.

### S1.4 Offline reproduction and optional live execution

To re-execute the eight final integration and reference trajectories from stored decisions, use scripts/reproduction/replay_saved_actions.py in the repository, as in Section S1.1. This route invokes neither a model nor an optimiser; it is distinct from archive-only hash and arithmetic checks.

The archived-response route executes the repository's three-role client through python -m examples.coordinated_planning.run --offline, using the ZIP directly as shown in Section S1.1. A3 has three periods; all selects A3, B3, C1 and C2 (159 periods, 483 saved responses). No separate client ZIP is required. Section S5.4 distinguishes current offline verification from historical live execution.

For new decisions, the optional live route requires credentials, verified token prices and an explicit spending limit; its command is in examples/coordinated_planning/README.md in the repository. The historical packaged client completed a three-period A3 live check on 18 November 2026. The repository migration was checked offline only; it does not constitute a new live or cross-platform check.

### S2. Related software and the basis for comparison

The comparison is purposive and based on primary documentation or original papers, with software citations checked on 8 November 2026 and the literature review extended on 14 November 2026 to include the original schlably and mango publications and supply-chain LLM-MAS studies. The main table covers environments with production, supply-chain or business-operation tasks. The dimensions are the reference or evaluation target, reported outcomes, and documented information or diagnostic support. Information parity requires examination of each experimental protocol. An unmentioned capability is not classified as absent. No competitor was executed for this paper, and no exhaustive or performance comparison is claimed.

InvAgent [5] studies LLM-based ordering in a multi-echelon inventory environment. Jannelli et al. [6] examine information sharing, tool use and negotiation in a three-echelon supply chain and describe an open-source implementation integrated into a sequential supply-chain simulation environment. They report cumulative costs and a bullwhip measure; profit is not reported. For both Gemini models tested, the lowest-bullwhip configuration did not have the lowest cumulative cost [6, main-text Table 2]. Factory-simulator reports final cash alongside horizon profit, which includes outstanding receivables and stock salvage value.

| Environment | Reference or evaluation target | Information, control and diagnostics |
|---|---|---|
| OR-Gym [10,11] | MILP and heuristics; inventory-task shrinking-horizon LP and perfect-information oracle; task rewards. | Inventory state and action history; task-specific optimisation models. |
| AIM-Bench [2] | Ex-post replenishment reference; ordering deviations and inventory costs. | Multi-agent settings and information-sharing treatments; period-level ordering behaviour. |
| InventoryBench [12,13] | Capped base-stock baseline; normalised reward for OR, LLM and hybrid methods. | Pregenerated demand and lead times; historical and contextual inputs; recommendation or forecast flows. |
| SupChain-Bench [23,24] | Ground-truth answers and retrieval targets; accuracy and tool-use costs. | Context and procedural-guidance conditions; execution-error analysis. |
| Vending-Bench [1] | Long-running business operation; final net worth and a single-run human baseline. | Agent trajectories and tool use; LLM-generated supplier replies and demand parameters; noisy sales. |
| schlably [16] | DRL agents compared with dispatching heuristics and a solver; reward and gap to the solver solution. | Generated scheduling instances, Gym environments, logged outcomes and Gantt charts. |
| SCML [17,18] | Comparisons of negotiating agents in supply-chain worlds. | Configurable contexts and controlled world conditions, including exogenous contracts and their revelation times. |

| Environment | Reference or evaluation target | Information, control and diagnostics |
| --- | --- | --- |
| MLPro-MPPS [19,20] | Reusable production environments for different learning methods; modular, decentralised Bulk Good System example. | Sensor and component states; states may be unavailable to sensors; integration with reinforcement learning and game theory. |
| prodsys [21] | Production configurations; throughput, throughput time, work in progress and reconfiguration cost. | Seeded simulations and stored events; reconfiguration cost of machines, transport resources and process modules as an optimisation objective or constraint. |
| SimPROCESD [22] | Maintenance and production policies; net production profit and quality. | Equipment condition and resource-use indicators. |
| EconEvals [14,15] | Procurement and pricing outcomes relative to OPT; scheduling blocking pairs relative to random matching. | Agent-neutral tool interface, trajectory analysis and litmus tests with conflicting objectives; note analysis assumes notes faithfully describe reasoning. |

Software references identify the associated repositories separately from the papers describing OR-Gym [10,11], InventoryBench [12,13], MLPro-MPPS [19,20], SupChain-Bench [23,24] and EconEvals [14,15]. SCML documentation and software are cited separately [17,18]; prodsys and SimPROCESD are identified by their software sources [21,22]. AIM-Bench and Vending-Bench are compared from their original papers [1,2]. Repository citations identify sources consulted; no competitor was executed. Living documentation may change after consultation.

The schlably framework integrates scheduling-instance generation, Gym environments and comparisons of deep reinforcement learning agents with dispatching heuristics and a solver [16, Sections 2–4]. Its illustrative evaluation reports scheduling outcomes and solver gaps. This is a precedent for reference-based production-decision evaluation. Factory-simulator focuses on financial consequences across procurement, production and fulfilment, with receivables and stock salvage, recorded-history forks and configurable disclosure. These are differences in the documented research scope, not evidence that schlably cannot be extended. mango supplies agent roles and communication infrastructure, including external-simulator coupling [25, Sections 3–4]. It provides complementary architectural context; integration with Factory-simulator has not been tested.

Publication status was checked on 14 November 2026. EconEvals [14] is listed among accepted EC 2026 papers; its coauthor's publication list distinguishes the conference abstract from the full working paper. The comparison uses the full arXiv v4 text. Sources: https://ec26.sigecom.org/program/accepted-papers/ and https://yannai.gonch.name/scientific/research.html. References [1,2,5,10,12] retain their arXiv citations because no replacement publication was verified.

InventoryBench applies compared methods to the same pregenerated demand and lead-time realisations [12]. Its normalised reward uses an optimistic bound, not a proven attainable instance optimum. Product context and dates are available to language-model procedures, while the base OR procedure does not use that text [12, Appendix E]. These comparisons study complementary procedures; they do not isolate algorithm effects under identical use of information. AIM-Bench reports similar stockout rates with different deviations from optimal ordering [2, Finding V]. This supports examining decisions beyond one operational measure, although the comparison does not concern profit or payment timing.

MLPro-MPPS describes reproducible simulations and training and reuse across learning methods, supporting controlled comparisons [19]. Component states may be unavailable to sensors, a precedent for restricted observation. Neither establishes a recorded-history branching mechanism. EconEvals analyses agent notes under an explicit assumption about their faithfulness [14]. Our recorded actions do not expose a controller's internal reasoning.

## S3. Scenario catalogue

The following catalogue describes all 23 supplied scenarios. Definitions come from docs/library/scenario-library.md in the repository and the supplied configurations, not from the illustrative micro-factories. All run for 52 periods. The common factory starts with no cash, about three periods of materials and components, and a quarter-period of finished stock. The credit limit is 80000. Initial distribution-centre stocks are 22.500 industrial controllers, 23.750 smart monitors and 20.000 sensor gateways. There are no initial supplier orders. Individual scenarios override this start where stated.

The suggested diagnostics identify mechanisms to inspect. They are not guaranteed directional effects. A study should define its controller, matched comparator and seed population before using these cases to make performance claims.

### S3.1 Baseline

#### 1. steady-market

Unchanged demand over 52 periods and no additional disruption.

Establish a common operating reference. Inspect fulfilment, profit, inventory and resource utilisation.

### S3.2 Demand

#### 2. seasonal-demand

Four 13-period seasons multiply demand by 0.2, 0.6, 6 and 0.4.

Examine whether procurement and production build stock before the peak without exhausting warehouse space. Track transfers and lost sales.

#### 3. growth

Demand grows by 5% per period, reaching approximately twelve times its starting level.

Investigate when increasing demand exceeds capacity or liquidity. Track lost sales alongside production and purchases.

#### 4. decline

Demand declines by 12% per period, reaching approximately 0.15% of its initial level.

Study when a controller reduces procurement and production. Track stranded inventory and terminal salvage.

#### 5. demand-surge

Smart-monitor demand becomes eight times its usual level in periods 14–41 without advance notice.

Inspect adaptation to a temporary product-specific peak, including missed orders and competing resource demands.

### 6. demand-step

All product demand becomes six times its usual level from period 11 onward.

Compare response to a permanent demand shift with response to a temporary surge. Track fulfilment and capacity.

### 7. product-life-cycle

Sensor gateways start without sensor modules or finished stock. An S-shaped demand profile rises to three times normal around period 6 and ends in period 20; steepness is 1.4.

Examine launch preparation and cessation of purchasing before obsolescence. Inspect gateway sales by demand phase and order cohort, alongside remaining inputs. Comparisons with steady-market also change initial stock.

## S3.3 Suppliers

### 8. supplier-outage

During periods 10–41 the Asian chip supplier delivers nothing, and the European supplier cannot sell microchip-b. No advance notice is given.

Trace different consequences of failed deliveries and blocked purchases. Inspect material availability and downstream fulfilment.

### 9. transport-delays

First supplier deliveries due in periods 12–41 are delayed by 16 periods. Backfill deliveries retain their own schedule.

Study pipeline exposure and delayed availability. Track late arrivals, open commitments and resulting lost sales.

### 10. quality-crisis

The acceptance rate of the European parts supplier is multiplied by 0.001 in periods 9–40. It uniquely supplies PCB blanks and connector pins.

Inspect how rejected inputs propagate through dependent recipes. Record accepted quantities, scrap and missed orders.

### 11. material-shortage

Maximum purchases of both chip types become zero in periods 10–41.

Study procurement restrictions separately from delivery failures. Track rejected purchases and inventory depletion.

### 12. price-change

Microchip purchases cost four times normal in periods 8–39 without notice. Microchip-b retains its price; it can substitute only in the control-board alternative recipe.

Examine recipe substitution and, where price changes are anticipated, advance purchasing under credit constraints. Track unit prices, purchase quantities, material choices and payment timing alongside profit. Total spending need not increase when a unit price rises.

### S3.4 Production lines

#### 13. line-breakdown

Final assembly loses all condition in period 18 and has no regular or overtime hours in periods 22–37.

Separate condition restoration from a period with no available hours. Inspect repairs, downtime and delivery losses.

#### 14. frequent-breakdowns

Final assembly faces an additional 80% per-period chance of losing 60–90% condition, and has no hours in periods 34–41.

Study repeated recovery decisions under severe disruption. Track maintenance resource use and lost sales.

#### 15. capacity-cut

The SMT line has no regular or overtime hours in periods 18–41.

Assess reliance on intermediate stocks and other processes during a sustained capacity interruption.

#### 16. yield-drop

SMT production started in periods 7–40 yields 95% fewer good units. A base yield of 0.975 becomes 0.04875.

Inspect resource consumption that produces little usable output. Track started units, good output and scrap separately for matched production cohorts. Report yield ratios only for positive production; zero production leaves them undefined.

### S3.5 Cash and space

#### 17. cash-squeeze

Initial cash is −120000, or 40000 below the credit line, with a quarter-period of finished goods.

Purchases remain inadmissible until customer payments restore sufficient cash after commitments. Study shipment priorities and recovery of purchasing ability.

#### 18. full-warehouse

The main plant is full, with stock value concentrated in 6347.380 enclosures, 820.615 LCD panels, 1554 connector pins and 2823 copper wire; other inputs are absent.

Cable-harness production can release space for missing inputs. Inspect space-releasing actions alongside on-time deliveries, late deliveries, lost sales and liquidity. Fewer late deliveries can accompany reduced fulfilment.

### S3.6 Information

#### 19. forecasts

Base demand is hidden. Forecasts cover six periods, with error up to 20% one period ahead and an additional bound of up to 4% per further period, reaching 40% at six periods.

Use a hidden-demand control without forecasts to isolate forecast provision. Comparing directly with steady-market also changes parameter visibility. Compare decisions and business outcomes on matched worlds; improvement depends on the controller’s use of forecasts.

#### 20. notice

This scenario is labelled “Outage with notice” in the article; its code identifier is notice.

The same chip-source outage as supplier-outage occurs in periods 10–41, but both restrictions are visible from period 6.

Compare against supplier-outage at the same seed to study four-period advance disclosure. Inspect purchasing changes and service; benefit is not assumed.

#### 21. hidden-groups

Demand is multiplied by 2.5. Demand parameters, supplier delivery and inspection parameters, line yield and breakdown parameters are hidden.

Use a counterpart retaining the 2.5 demand multiplier with parameters shown. A comparison with steady-market confounds demand and visibility.

### S3.7 Combined disturbances

#### 22. regional-crisis

Asian supplier deliveries arriving in periods 16–41 are lost. First deliveries due during the crisis are delayed by 16 periods.

Inspect how delivery loss and delay interact. Separate committed cash from usable material and eventual fulfilment.

#### 23. surge-and-breakdown

Smart-monitor demand becomes eight times normal in periods 14–41, while the SMT line loses all condition in period 16.

Study coordination of capacity recovery with demand response. Compare component shortages, repairs and customer prioritisation.

The release documentation reports a development campaign of 520 worlds and 3948 runs, including seed-0 repeats, covering 23 scenarios and three controls on seeds 0–19. The release documentation reports no violations of its run-level checks, including material and financial balances, across the 3948 development runs. Four worlds were unscorable because the reference optimiser found no solution within its limits. These run-level checks do not establish whether the expected outcome direction holds. The documentation lists six failed direction checks: decline, product-life-cycle, price-change, full-warehouse, forecasts and yield-drop. Each check compares a designated outcome with steady-market for the same controller and seed. It requires a strict decrease for decline and product-life-cycle under the full-horizon, rolling and aligned optimisers; a strict increase for price-change, full-warehouse and yield-drop under those optimisers and the full-foresight reference; and a strict decrease for forecasts under the rolling optimiser. A tie or an opposite change in any available controller–seed pair fails the direction check. These statements describe controller outcomes, not tests of whether a configured mechanism executes correctly. Controller variants are defined in Section S4.

The separate difficulty assessment uses the buffered rolling optimiser and paired calibration-score differences across seeds. Unlike Equation (S1), this score uses the steady-market reference-minus-no-decision profit difference for the same seed as a common denominator. It equals one minus the scenario-specific reference-to-controller profit gap divided by that denominator. The code withholds a difficulty verdict for scenarios classified as involving foresight. The documentation reports difficulty as yes for yield-drop despite its failed direction check. Thus, the six failed direction checks do not identify six scenarios below a common difficulty threshold. The available release summary does not identify the individual difficulty outcomes for decline, product-life-cycle, full-warehouse and forecasts; no outcomes are inferred for them here. For price-change, the code assigns the foresight category rather than a difficulty verdict.

The scenario catalogue provides configurable business conditions. Its suggested diagnostics should be selected for the research question, controller and information regime. A study claiming calibrated difficulty requires a separately specified calibration protocol. The available development summary does not contain the historical failing controller–seed rows; their individual causes therefore remain unresolved. The software's development campaign was not repeated for this article. Definitions and reported results are in tests/stress/playing.py, tests/stress/report.py and docs/benchmark/stress-test.md in the repository. Selected mechanism checks for product-life-cycle, price-change, yield-drop, full-warehouse and forecasts appear in tests/integration/examples/test_scenario_library.py. Their coverage should not be equated with complete scenario validation or difficulty calibration. For decline, the scenario-library integration tests contain no dedicated demand-decline mechanism check; generic tests cover its configuration and API registration.

### S4. Optimisation references and information boundaries

| Controller | Available information | Planning role |
|---|---|---|
| No decisions<br>no_decisions | No decision input required | Empty action list; zero-score anchor. |
| Full-foresight reference<br>reference_optimizer | Generated world and customer orders across the horizon | Repeated full-horizon LP/MILP, followed by engine checks. |
| Full-horizon limited<br>full_horizon_optimizer | Description and current observations | Expected-value plan through the remaining horizon. |
| Rolling<br>rolling_optimizer | Description and current observations | Window based on the longest buy-to-sale lead-time chain plus two periods. |
| Buffered rolling<br>buffered_rolling_optimizer | Same as rolling | Terminal stock multiplier 1.2 instead of 1.01. |
| Forecast-aligned<br>aligned_optimizer | Description and current observations | Rolling window; demand from open orders and forecasts. |
| Myopic<br>myopic_optimizer | Description and current observations | Current-period plan with terminal inventory and pipeline valuation. |
| Informed diagnostic<br>informed_optimizer | Observations plus means of otherwise hidden parameters | Restores parameter means from an unhidden counterpart; does not reveal future draws. |

| Controller | Available information | Planning role |
|---|---|---|
| Exact-integer reference control reference_optimizer(..., exact=True) | Same full foresight as the reference | Integer choices in every modelled period; limited to small programs of at most 100 binary choices. |

The ordinary reference relaxes binary choices beyond the first period of each program. Its mathematical solution is rounded and translated into ordered actions. Feasibility checks may fit quantities or drop actions. The model and execution need not coincide, so a small solver gap is not an engine-level optimality certificate. The exact-integer control removes one relaxation but does not establish that every engine mechanism is represented exactly.

For a planning window ending before the run horizon, rolling and buffered rolling value eligible terminal stock at 1.01 and 1.20 times replacement cost, respectively, up to the expected-use quantities computed beyond the window. Full-horizon limited plans through the remaining run horizon. At the run horizon, the manifest's salvage values apply instead of the intermediate-window valuation. Here, limited denotes the use of descriptions and current observations rather than the realised future world.

For the controlled demonstrations in Sections S5.1–S5.3, the time limit was 30 seconds per solve. The implementation also imposes a 1000-node limit and an absolute gap stopping tolerance of 0.01 monetary units. Recorded solves found solutions and did not stop at the time or node limits. The revised notice matrix has reported gaps no larger than $4.55 \times 10^{-13}$. The inspected optimiser runs reported no fitted or dropped actions. Section S5.2 specifies the tighter tolerance used for its separate planning diagnostics.

Limited controllers see disclosed parameters and observations, including visible disruption duration in the notice example. Administrative world endpoints are not an information-safe interface. Experiments must restrict access in their harness or deployment. The repository includes demonstration clients for single-prompt, tool-guided and coordinated planning; study configuration remains the researcher’s responsibility. Deterministic traces expose engine execution, not a model’s private reasoning. A language-model harness should also record model configuration, prompts and every action, distinguish replay of stored actions from new sampling, and report controller compute or token budgets where relevant.

### S4.1 Specifying a decision-evaluation study

The following protocol describes how the existing engine and controller interfaces can support a future study. It is not a report of an LLM or MAS experiment. Define the objective and horizon before inspecting comparative results. Specify terminal valuation and each service measure, including whether its denominator counts orders, units or value, and whether it covers orders received, orders due or completed shipments. Report the physical or financial units. The notice demonstration counts 40 monitor units demanded and shipped on time; it is not a count of 40 distinct orders. In Vending-Bench, tools advance simulated time by different amounts and the message cap can produce different elapsed horizons [1]. A study should therefore distinguish decision counts, simulated duration and compute budgets.

For a controller comparison, hold the factory specification, scenario and exogenous world fixed. Use a defined set of seeds for inference over worlds. State the disclosed parameters, forecast availability and permitted endpoints. Observation-limited comparators obtain the same type of description and observations as the tested controller, with equivalent opportunities to check feasibility. Their trajectories can nevertheless produce different observations. The comparison does not require identical realised state after different actions.

Report comparator planning windows, assumptions and solver budgets. A full-horizon observation-limited comparator estimates future conditions from disclosed information; it does not see the realised future. The informed diagnostic restores hidden distribution means. The full-foresight reference receives the realised world and future customer orders. Its outcome is diagnostic, not a guarantee of the optimum attainable under the tested controller's information. State these distinctions alongside any normalised score.

An information intervention holds the decision procedure and physical world fixed while changing disclosure. A comparison of different algorithms is a different contrast. To study both factors, specify matched conditions for each and analyse their interaction where relevant. Subtracting a limited controller's profit from a full-foresight reference does not by itself separate information effects from planning or execution effects.

For a fork, specify the intervention period and which actions change. Restore or deliberately redefine the external controller's memory and communication state; the engine reproduces factory history, not arbitrary controller state. State whether subsequent actions are replayed or generated again. Inspect rejected actions and material balances alongside the financial record. A resulting difference describes that continuation under the model; it does not reveal a language model's internal reasoning or establish a policy ranking across worlds.

Archive controller configuration and submitted actions. Stochastic controllers require repeated runs or another justified uncertainty treatment in addition to matched worlds. Compute budgets and API costs should be reported separately or incorporated into a declared study objective; the engine's factory profit does not automatically charge an external LLM service. Record solver statuses and score exclusions. The demonstrated pipeline provides execution evidence; industrial use requires validation for the intended factory and decision purpose.

### S4.2 Financial interpretation and terminal conditions

Horizon profit is the model-defined financial outcome in Equation (1). Initial cash is subtracted; the acquisition value of initial inventory is not charged. Stock remaining in warehouses is valued at its specified salvage rate. Work in progress and goods in transit receive no terminal salvage value; costs already incurred remain charged. Outstanding customer receivables count at their full nominal value, including payments due after the horizon. Undelivered supplier quantities are unpaid and carry no terminal asset or liability value. At final close, every open customer order whose due period has arrived incurs the lost-sale penalty for its unshipped quantity; orders not yet due earn nothing and incur no terminal penalty. Earlier costs and penalties remain in the ledger. Comparisons across controllers should hold initial conditions and valuation rules fixed. If $V_0$ is a common valuation of initial warehouse stock and $\pi' = \pi - V_0$, then $\pi'_1 - \pi'_2 = \pi_1 - \pi_2$. Subtracting $V_0$ from the controller, no-decision baseline and reference also leaves Equation (S1) unchanged. Absolute outcomes shift, so the reported horizon profit must be interpreted using the stated endowment convention. Comparisons across scenarios require accounting for differences in initial endowments, horizons and valuation assumptions. Terminal conventions also differ across benchmarks. OR-Gym assigns no terminal value to remaining inventory in its inventory task [10]; InventoryBench sums sales margins less holding costs without a separate terminal-stock valuation term in the stated objective [12]. Vending-Bench includes stock at wholesale value in final net worth [1]. Their numerical financial scores are therefore not directly interchangeable.

The no-decision baseline $\pi_0$ and full-foresight reference $\pi_r$ define the diagnostic score

$$s = (\pi - \pi_0) / (\pi_r - \pi_0). \qquad \text{(S1)}$$

When $\pi_r > \pi_0$, zero denotes baseline profit and one denotes reference profit; otherwise the score is undefined. Scores can exceed one because the reference is heuristic. The score is neither an optimality percentage nor an information-value measure.

### S5. Worked examples and numerical evidence

S5.2 supports main-text Section 3.1 (advance notice); S5.1 supports Section 3.2 (alternative assembly); S5.3 supports Section 3.3 (checkable optimum); S5.4 and S5.5 support Section 3.5 (live LLM-MAS integration and matched-world references). All distributed demonstration records were generated against the identified 1.0.0 snapshot.

### S5.1 Order fulfilment and fork

The first example uses the plain tiny factory with no recurring demand, initially empty stock and cash 1000. A four-monitor order is due in period 4 with one period of patience and one-period payment terms. Each board requires two chips and each monitor two boards. Unit costs are chip 0.5, board production 1, monitor production 2 and shipment 1; monitor price is 50. The late-delivery penalty is 2 per unit per period, and the lost-sale penalty is 10 per unit. The delayed shipment incurs 4 × 2 = 8; the no-decision run incurs that penalty plus 4 × 10 = 40 at final close. Line capacity is ten hours per period, with 0.1 hour per board and 0.5 hour per monitor. Supplier and production lead times are one; yield is one, and wear, storage charges and interest are zero.

| **Outcome** | **Original schedule** | **Delayed fork** | **No decisions** |
|---|---|---|---|
| Shipment period | 4 | 5 | None |
| Final cash | 1172 | 964 | 952 |
| Receivables | 0 | 200 | 0 |
| Penalties | 0 | 8 | 48 |
| Profit | 172 | 164 | −48 |

The fork at period 3 preserves periods 1–2 and moves monitor assembly from period 3 to period 4. The shipping rule remains unchanged. Equality checks cover the closed prefix, a repeated parent, replay followed by continuation, and the fork against a fresh run of the same alternative. Hiding line yield changes its presence in the controller description but leaves physical records unchanged for this fixed policy.

Seven runs passed aggregate material and financial reconciliation. The prefix digest is f5400475079e261332723e0c7279f64528fe6671b136a3ebba48d57e572e4097. The full parent, repeated-parent and replay-continuation digest is b0354f7459b7dd998726c4992550b8eec89e7b506f3882c65de29c43db00d6a1. Serialisation is json.dumps(value, default=str, sort_keys=True, indent=2), encoded as UTF-8. Prefix hashes cover the list of two period rows; full hashes cover the period-response object. The procedure is scripts/reproduction/article/run_demonstrations.py in the repository; records are in examples/order-fulfilment/ in the data archive.

### S5.2 Advance notice

The ten-period factory, seed and event schedule are specified in Section 3.1. The purchase blockage in periods 3–6 is disclosed either at onset or in period 1. The main example uses storage charges of 0.0023 per chip, 0.0090 per board and 0.0270 per monitor per period. These charges are approximately 0.45% of replacement costs (0.46% for chips); they are constructed inputs, not industrial calibration estimates. The library-like label refers to storage rates around 0.41–0.51% of replacement cost documented in docs/benchmark/optimizers.md in the repository; the example does not reproduce all library-factory parameters. Matched worlds agree after removing only the disruption visibility date. All 34 matrix runs and the additional controlled continuation passed material and financial reconciliation, and all submitted decisions were accepted.

*Table S5.2a. Positive storage charges. Cells report on-time deliveries out of 40 and profit in monetary units.*

| Controller | Disclosure at onset | Notice in period 1 |
|---|---|---|
| No decisions | 0/40; -388.5680 | 0/40; -388.5680 |
| Rolling | 24/40; 919.1720 | 28/40; 1130.9552 |
| Buffered rolling | 24/40; 919.1720 | 40/40; 1732.4280 |
| Full-horizon limited | 24/40; 919.1720 | 40/40; 1766.0840 |
| Full-foresight reference | 40/40; 1766.0840 | 40/40; 1766.0840 |

With advance notice and positive storage charges, buffered rolling and full-horizon limited both ship all 40 monitors on time. Buffered rolling incurs material payments of 96 and storage charges of 3.572, with terminal salvage of 8. Full-horizon limited incurs material payments of 56 and storage charges of 1.916, with zero salvage. Revenue, production and transport costs coincide. The profit difference is therefore 40 + 1.656 − 8 = 33.656. Their policies differ in both planning window and terminal inventory valuation, as specified in Section S4. Full-horizon limited and the full-foresight reference obtain the same realised outcome in this instance.

*Table S5.2b. Zero storage charges. Cells report on-time deliveries out of 40 and profit in monetary units.*

| Controller | Disclosure at onset | Notice in period 1 |
|---|---|---|
| No decisions | 0/40; -386.4000 | 0/40; -386.4000 |
| Rolling | 40/40; 1768.0000 | 40/40; 1719.2000 |
| Buffered rolling | 40/40; 1768.0000 | 40/40; 1719.2000 |
| Full-horizon limited | 24/40; 920.0000 | 40/40; 1768.0000 |
| Full-foresight reference | 40/40; 1768.0000 | 40/40; 1768.0000 |

The archived zero-storage-cost runs were reproduced exactly after normalising JSON values, including period records, final results and generated worlds, for no decisions, rolling and the reference in both disclosure regimes. Rolling earns 1768 without advance notice and 1719.2 with notice, shipping all 40 monitors in each case. In the no-notice run it buys 80 chips in period 2, before the blockage is observable. Re-solving the recorded period-2 planning problem with purchases fixed to either 16 or 80 gives the same objective, 1081.44. Replacing only this purchase with 16, keeping the other period-2 actions and subsequent rolling policy unchanged, reduces timely deliveries to 24 and profit to 920. The zero-cost rolling result therefore depends on selection among equally valued planning solutions.

A positive charge on chips alone (0.0001 per unit per period) changes the returned no-notice schedule but does not eliminate the planning tie: chips can be transformed into boards with zero storage charge. With the positive charges used in Section 3.1, the recorded no-notice period-2 program instead values a purchase of 16 at 1079.856 and a purchase of 80 at 1079.5984. This establishes a strict preference between those two choices in that program, not uniqueness of the complete plan.

*Table S5.2c. Rolling sensitivity. Rates are percentages of replacement cost per period unless stated otherwise; cells report on-time deliveries out of 40 and profit.*

| Storage rate | Charges: chip / board / monitor | Disclosure at onset | Notice in period 1 |
|---|---|---|---|
| Zero | 0 / 0 / 0 | 40/40; 1768.0000 | 40/40; 1719.2000 |
| Chip only: 0.0001 | 0.0001 / 0 / 0 | 24/40; 920.0000 | 40/40; 1719.1744 |
| 0.02% | 0.0001 / 0.0004 / 0.0012 | 24/40; 919.9632 | 40/40; 1735.8432 |
| 0.1% | 0.0005 / 0.002 / 0.006 | 24/40; 919.8160 | 40/40; 1735.2160 |
| 0.2% | 0.001 / 0.004 / 0.012 | 24/40; 919.6320 | 40/40; 1734.4320 |
| About 0.45% | 0.0023 / 0.0090 / 0.0270 | 24/40; 919.1720 | 28/40; 1130.9552 |
| 1% | 0.005 / 0.02 / 0.06 | 24/40; 918.1600 | 28/40; 1129.6800 |
| 2% | 0.01 / 0.04 / 0.12 | 24/40; 916.3200 | 28/40; 1127.3600 |
| 5% | 0.025 / 0.1 / 0.3 | 24/40; 910.8000 | 28/40; 1120.4000 |

The rolling sensitivity sweep uses storage charges of 0.02%, 0.1%, 0.2%, 1%, 2% and 5% of replacement cost, alongside the approximately 0.45% main case. Without advance notice, all tested positive-charge cases deliver 24 monitors. With notice, the tested 0.02%, 0.1% and 0.2% cases deliver 40; the main case and the tested 1%, 2% and 5% cases deliver 28. These sampled rates do not locate a general threshold. At 0.2%, changing only the terminal valuation multiplier from 1.01 to 1.00 in the recorded period-2 planning problem reverses its preference between purchases of 176 and 32. The objective differences, in favour of 176, are +0.672 and −0.768, respectively. This diagnostic concerns that fixed planning state; it is not a full-run ablation of the multiplier.

Six recorded planning inputs were reconstructed exactly. The original programs reproduced the recorded objective and current purchase before diagnostic constraints were introduced. The 42 planning solves preserve the release program's binary structure. Diagnostic comparisons use a relative gap target of zero, an absolute gap target of 1e-9 and a comparison tolerance of 1e-7, with constraint and integrality residual checks. They establish properties of the planning programs and the specified continuation, not general engine-level optimality. Source 1.0.0 was unchanged. Full records, input manifests and solver reports are provided in the data archive's notice/ directory. Reproduction scripts are in the code repository under scripts/reproduction/article/ (Section S1.3).

Twelve additional runs extend the advance-notice sensitivity assessment to buffered rolling and full-horizon limited at 0.02%, 0.1%, 0.2%, 1%, 2% and 5% of replacement cost per period. They use seed 7, ten periods and 30-second solve limits. All completed with accepted actions and reconciled material and financial balances; their worlds matched each other and the corresponding rolling run. Together with zero and the library-like setting from the repeated original matrix, both controllers deliver all 40 monitors at eight tested storage settings. In this constructed instance, their profit difference ranges from 32.07 to 50.00, calculated before rounding. These sampled settings do not establish the result throughout a continuous interval or across other factories.

Table S5.2d. Advance-notice sensitivity of the equal-service comparison. Both controllers deliver 40 of 40 monitors on time in every row. Rates are percentages of replacement cost per period; the library-like row uses charges 0.0023, 0.0090 and 0.0270 and is approximately 0.45%. Profit differences use unrounded values.

| Storage setting (%) | Buffered rolling profit | Full-horizon limited profit | Profit difference |
|---|---|---|---|
| 0 | 1719.20 | 1768.00 | 48.80 |
| 0.02 | 1735.84 | 1767.92 | 32.07 |
| 0.1 | 1735.22 | 1767.58 | 32.36 |
| 0.2 | 1734.43 | 1767.15 | 32.72 |
| ≈0.45 (library-like) | 1732.43 | 1766.08 | 33.66 |
| 1 | 1728.16 | 1763.76 | 35.60 |
| 2 | 1720.32 | 1759.52 | 39.20 |
| 5 | 1696.80 | 1746.80 | 50.00 |

### S5.3 Independently checkable optimum

The one-period micro-instance starts with 0.001 monitors and two due orders of 0.001 each. Prices are 50 and 70; shipping cost is 1 and unserved quantity incurs a penalty of 10 per unit. There is no recurring demand, interest or terminal salvage. Cash and credit are zero. Positive stocks of 0.001 chips and boards satisfy configuration validation but cannot become saleable monitors before shipment decisions. Production completes after the shipping phase.

The API permits quantity precision of at most three decimal places. Validation accepts 0.001 and rejects 0.0005, as recorded in historical/v06-verification/environment.json. Therefore the nonzero stock can be assigned only to the lower-price or higher-price order; the remaining alternative is no shipment. The three engine outcomes are −0.020, 0.039 and 0.059. There is no initial receivable, supplier order or other income source. Repairs are cost-neutral here and other actions cannot improve same-period revenue.

More generally, let $x$ and $y$ be shipments to the lower-price and higher-price customers, with $x + y \leq 0.001$. Profit from shipping and lost sales is $49x + 69y - 10(0.002 - x - y) = 59x + 79y - 0.020$. This is at most $79 \times 0.001 - 0.020 = 0.059$. Any additional positive cost only lowers profit. Shipping all stock to the higher-price order attains this bound. The proof thus also bounds a hypothetical continuous relaxation.

The ordinary reference, exact-integer reference and rolling controller each achieve 0.059 in engine execution. Seven runs, including the baseline and enumerated alternatives, pass financial reconciliation. The no-shipment policy is executed both as the no-decision baseline and as an enumerated alternative. This explains the run count. Results are in examples/reference/micro-*.json. This is a sanity check on a deliberately small case, not validation of reference optimality on the scenario library.

### S5.4 Live LLM-MAS integration

### Purpose and configuration

The final demonstration comprises one three-period small-factory run (A3), one 52-period steady-market run (B3), and two 52-period notice runs (C1 and C2). All use simulator seed 7 and fresh controller memory. A3 uses the repository's example fixture; B3 and both C runs use the distributed electronics-factory manifest and the steady-market or notice scenario. C1 and C2 have exactly matching world and description records. Scenario labels denote business conditions, not calibrated difficulty levels. These are integration cases, not a comparative effectiveness experiment.

The engine is the unmodified 1.0.0 source snapshot identified in Section S1. The historical demonstration client extended the repository’s coordinated example; the current repository contains its integrated implementation. Its planner proposes a plan, procurement reviews purchasing, and reviewer uses feasibility feedback to revise the decision list. The same hosted model serves each role sequentially; this is a role-separated orchestration, not a comparison of independently trained models. The client receives the disclosed description, current observation and summaries of the previous two closed periods, including cash, stock and executed outcomes. Administrative world records are retrieved for auditing after execution, not supplied as additional model inputs. Feasibility checking is available to the client; the service itself does not enforce information isolation.

Live calls on 12–13 November 2026 used OpenAI gpt-6.1-terra, reasoning_effort=low, service_tier=default and max_completion_tokens=8192. The recorded returned model identifier matched the requested identifier. No immutable provider snapshot suffix was available in these records, so the call date and returned identifier are retained. The per-request input limit was 16384 whitespace-separated words. It is not a token count. The final per-period allowance was five calls, comprising three role calls and up to two additional reviewer revisions. The client maintained an empty initial memory for each run. Full request bodies preserve prompts, role contexts and settings; response records preserve usage and returned text. Live provider resampling is not expected to reproduce the archived decisions.

### Historical client changes

Relative to the public example, the local adapter applies the selected model and an 8192-token output limit, and the client adds up to two further reviewer revisions when the proposed final plan remains infeasible. The additional instruction requests revision of the complete ordered decision array against the unchanged observation and supplied validation results. It explicitly addresses fixed setup and batch constraints as well as cash, credit, materials, storage and line capacity. If both revisions fail, a deterministic filter removes currently rejected actions and rechecks the remaining ordered plan; unresolved prior scheduled production causes a stop. No filter was invoked in the final four runs. Both C runs needed three additional reviewer calls, each resolving a different period; no period required both extra revisions.

The local transport permits at most two attempts per logical call, with connect/read/write/pool timeouts of 10/240/20/5 seconds and a 600-second logical deadline including waits. Retry eligibility is restricted to specified transient failures, with bounded waiting and persistent budget reservations; ambiguous partial responses stop execution. Simulator decision submissions are not blindly retried. The final series used no transport retries or timeouts, so it supplies no empirical estimate of retry effectiveness. Offline tests, rather than these live runs, exercise those branches. The settings file retains inherited adapter fields, including timeout_seconds=120 and retries=0; the separately recorded transport_policy specifies the actual local v3 transport limits.

### Observed results and checks

All four runs completed: 159 periods and 483 model responses, including six additional reviewer revisions. Every period and final profit passed cash, customer, supplier and item-balance reconciliation; no final executed decision was rejected. Role handoffs and revision rules were consistent. These checks reconcile recorded flows rather than independently verify every engine cost coefficient. Saved-action replay matched period records, final results and worlds (Section S5.5); it does not reproduce stochastic model decisions. Table S5.4a distinguishes the evidence types.

Financial values below are model units; shipment rates use unit denominators and are not order-count rates. Outstanding orders need not be either shipped or lost before the horizon.

A3: horizon profit 328; initial cash 1000; final cash 1328; outstanding receivables 0; stock salvage 0. Units ordered 12, shipped 8, shipped on time 8, shipped late 0, lost 4. Shipped/ordered 66.67%; on-time/shipped 100.00%; on-time/ordered 66.67%.

B3: horizon profit 552071.4520155; initial cash 0; final cash 361309.7147587; outstanding receivables 181597.1341; stock salvage 9164.6031568. Units ordered 14398.221, shipped 11281.293, shipped on time 3125.294, shipped late 8155.999, lost 2749.433. Shipped/ordered 78.35%; on-time/shipped 27.70%; on-time/ordered 21.71%.

C1: horizon profit 203626.7626826; initial cash 0; final cash -15502.6834614; outstanding receivables 211964.414; stock salvage 7165.032144. Units ordered 14398.221, shipped 7871.577, shipped on time 2607.455, shipped late 5264.122, lost 6198.043. Shipped/ordered 54.67%; on-time/shipped 33.12%; on-time/ordered 18.11%.

C2: horizon profit -7944.5234898; initial cash 0; final cash -119243.3940266; outstanding receivables 94956.9375; stock salvage 16341.9330368. Units ordered 14398.221, shipped 6053.889, shipped on time 2177.267, shipped late 3876.622, lost 7976.837. Shipped/ordered 42.05%; on-time/shipped 35.96%; on-time/ordered 15.12%.

A structural audit covered all 483 recorded logical model requests (A3: 9; B3: 156; C1 and C2: 159 each). Each supplied description matched the run-start record, and each observation matched the engine's record for that period. LLM-MAS and rolling runs on the same world received identical start descriptions. Parsed JSON message payloads contained no seed or world fields and no customer order with an arrival period later than the request period. In both C runs, disruption notices first appeared in period 6; the disruptions last 32 periods, from period 10 through period 41. This checks the recorded information path, not general access isolation or complete information equivalence between controllers. The repository’s scripts/reproduction/check_recorded_inputs.py reproduces these checks without model calls or engine execution; its current report is validation/repository-2026-11-19/input-audit.json.

C1 uses an additional reviewer revision in periods 6 (credit), 12 (available line hours) and 40 (quantity); C2 does so in periods 2 and 6 (credit) and 25 (quantity). C1 chooses empty plans in periods 29 and 31–34; C2 in periods 26, 29–33 and 51–52. These plans were generated by the model and were not forced by the filter. Their presence alone does not establish intent or decision quality. C2's cash falls below minus 80000, but the credit rule limits purchase admission rather than every subsequent debit (main text Section 2.2). The library factory applies per-period interest rates of 0.0003 to positive cash, 0.0015 to debt within the 80000 credit limit, and 0.004 to excess debt. C2 paid 3783.1685 in interest and earned 527.2157. Its first crossing below the limit occurred in period 6 when a due material payment of 11278.3829228 reduced cash to -80458.9960183 during delivery receipt. In period 52, customer payments of 23420.1615 were outweighed by lost-sale penalties of 25721.51, storage of 277.5305636 and interest of 275.8701, leaving final cash of -119243.3940266. The credit-admission check therefore does not guarantee a lower bound on all subsequent cash balances. Receivables are valued nominally; the zero-interest example in Section S5.1 deliberately isolates payment timing from financing costs.

The low reasoning setting was selected before execution to keep this integration demonstration within its monetary budget. It was not tuned for best controller performance, and these runs do not compare reasoning settings.

### Run selection and offline checks

All four final runs are reported, including negative-profit C2. Earlier configurations and interruptions remain in the archive and Section S7.2; the development sequence does not estimate an unchanged client's completion probability.

Historical-client checks comprised 272 passing client/transport checks, five reporting-bug regression checks and 18 continuation-guard checks; they are not added to the 3430 simulator tests.

### Historical client verification and repository integration

The earlier packaged client preserved historical prompts and decision rules while adapting paths, imports and budget initialisation. Its 15 November verification on macOS arm64 used Python 3.14.10, NumPy 2.5.3, SciPy 1.18.1 and uv 0.13.7 in a new virtual environment with an existing dependency cache. With sockets blocked, four cases consumed 483 archived responses over 159 periods; reconstructed requests, decisions, periods, results and worlds matched, without balance violations. Its 213 selected tests and two credential-gate tests are version-specific, overlapping earlier selections rather than additional simulator tests. Reports are retained in validation/client-offline-2026-11-18/. The revised data ZIP contains no client code; use the repository commit in Section S1. Checks of historical reproduction commit 394f68c on 19 November passed 148 existing selected tests plus four new path tests, and again matched all four offline cases and eight saved-action replays. Reports are in validation/repository-2026-11-19/. No provider calls or full simulator/stress suite were repeated. Other-platform historical checks do not validate the current repository client’s live launcher.

On 18 November 2026, the freshly extracted packaged client completed one three-period A3 live check on macOS arm64 in a new virtual environment with an existing dependency cache: nine gpt-6.1-terra responses at low reasoning effort, profit 328 model monetary units, no retries, extra revisions, filtering or balance violations, at an estimated USD 0.04. Records, exact cost calculation and comparison with historical A3 are in validation/client-live-2026-11-18/; this check does not replace Table 3's A3 or establish stochastic reproducibility, 52-period completion or cross-platform operation.

*Table S5.4a. Integration evidence and its scope. Counts describe different checks and must not be summed.*

| Evidence | Scope and interpretation | Records / reproduction |
|---|---|---|
| Live demonstrations | A3, B3, C1 and C2: 159 periods, 483 responses; demonstrates execution, not a controller ranking. | integration/2026-11-12/ and integration/2026-11-13/; historical/integration/; Section S5.4 |
| Selected code tests | 213 selected tests plus two credential-gate tests for other examples; includes mocked failure handling. | validation/client-offline-2026-11-18/ (historical selection) |
| Recorded-input audit | 483 requests checked against descriptions and observations; evidence of the recorded information path. | scripts/reproduction/ check_recorded_inputs.py in the repository; validation/repository-2026-11-19/input-audit.json |
| Saved-action replay | Eight trajectories, 367 periods, including four optimisation references; checks engine execution of stored decisions. | Repository saved-action utility; validation/repository-2026-11-19/replay-check/; Section S5.5 |
| Archived-response execution | Four LLM-MAS trajectories, 159 periods; checks orchestration, reconstructed requests, decisions and engine outputs. | validation/repository-2026-11-19/ client-check/; repository client README.md |
| Packaged live check | One A3 run, 3 periods, 9 requests, 18 November 2026; exercises the live path, not 52-period or cross-platform operation. | validation/client-live-2026-11-18/ |

### Data, costs and reproduction

The data archive contains prompts, configurations, decisions, worlds, periods, request/response logs, audits and reports, without credentials or executable code. Final runs are in integration/2026-11-12/ and integration/2026-11-13/; earlier attempts are in historical/integration/. Code is supplied exclusively through the repository. Section S1 identifies the published code, data release and archive checksum. New hosted-model calls require credentials and can incur charges. Archive integrity checks and engine replay remain distinct; the repository README gives the commands.

API costs are separate from model-unit factory profit. Rounded usage-based estimates are USD 0.02 for A3, 4.48 for B3, 4.15 for C1 and 3.81 for C2; the total calculated before rounding is approximately USD 12.47. Successful responses across development series total approximately USD 15.89. Exact usage, rates and calculations remain in historical/integration/; the conservative budget ledger includes uncertainty from the B2 timeout. These estimates are not reconciled invoices. Model-unit profits exclude them, so these runs do not establish net economic benefit after inference and coordination costs.

### S5.5 Matched-world optimisation references for the integration demonstration

Four additional reference trajectories were specified after the LLM integration runs to contextualise their outcomes: rolling and full foresight on steady-market and outage-with-notice, seed 7, each for 52 periods. Each uses the unchanged 1.0.0 release. The observation-limited rolling controller receives the disclosed description, observation and feasibility-check responses. Full foresight receives all customer orders and the realised future world from the matching archived no-decision run; exact-integer mode is disabled. It is an attainable diagnostic under additional information, not a certified engine optimum. Archived no-decision and myopic outcomes are retained as context, rather than selected as the only comparators.

The archived steady-market and outage-with-notice worlds differ only in disruptions. The outage-with-notice world adds an apac-chips supplier outage and a microchip-b material shortage, both starting in period 10, lasting 32 periods and disclosed from period 6. Customer demand and all other world fields are identical. Comparing these worlds does not isolate the value of advance information. That would require a matched supplier-outage control at the same seed, which was not run for this demonstration.

The protocol allowed 30 seconds per solve and 5400 seconds per reference trajectory, including recording and verification. Incomplete runs would receive no horizon profit or score, with no automatic budget extension; all four completed within this protocol. Execution used an interactive macOS arm64 workstation, with at most two optimisers and briefly one saved-action replay running concurrently. Machine load was recorded. The timings are not an isolated runtime benchmark, and fresh time-bounded optimisation may depend on computational load.

The completed reference runs exercised service routes through the in-process ASGI test client, with external network connections disabled. They incurred no model calls or API charges.

steady-market, rolling: profit 868970.31344082; final cash 679674.99717842; on-time/ordered 65.85%; score 0.872800. 52 period-level solve reports, 52 usable solutions, 0 time-limit and 0 node-limit stops; maximum reported absolute program gap 0.0019675466464832425 model units. Status codes [0]; fixed-binary LP status codes [0]. Total execution including audit and recorded-action replay 140.44 seconds.

In the steady-market world, orders arrive one or two periods before their due dates. Rolling combines open orders with expected future demand; in its run, late shipments were recorded in 40 periods, from period 7 to period 52. The cause of the on-time difference from full foresight was not tested. The order and shipment facts come from the archived records; the expected-demand planning source is set in the release code.

steady-market, reference: profit 1098988.810187; final cash 848010.4440806; on-time/ordered 99.58%; score 1.000000. 52 period-level solve reports, 52 usable solutions, 0 time-limit and 0 node-limit stops; maximum reported absolute program gap 6.430782377719879e-05 model units. Status codes [0]; fixed-binary LP status codes [0]. Total execution including audit and recorded-action replay 548.35 seconds.

Outage with notice, rolling: profit 179458.79328454; final cash -35284.99310586; on-time/ordered 36.74%; score 0.554937. 52 period-level solve reports, 52 usable solutions, 0 time-limit and 0 node-limit stops; maximum reported absolute program gap 0.0013620160752907395 model units. Status codes [0]; fixed-binary LP status codes [0]. Total execution including audit and recorded-action replay 100.60 seconds.

Outage with notice, reference: profit 892273.6753527; final cash 627924.9001535; on-time/ordered 82.99%; score 1.000000. 52 period-level solve reports, 52 usable solutions, 2 time-limit and 0 node-limit stops; maximum reported absolute program gap 59.132628857158124 model units. Status codes [0, 1]; fixed-binary LP status codes [0, 1]. Total execution including audit and recorded-action replay 511.53 seconds.

In notice-reference period 1 the MILP reported an optimal program solution, but the subsequent fixed-binary LP reached its remaining time budget. In period 2 both the MILP and fixed-binary LP reached their respective budgets. The unchanged solver retained the available MILP incumbent in each case, as implemented by its fixed-LP timeout fallback. The period-2 gap of 59.132628857158124 is an absolute program-objective gap, not a percentage or an engine-level gap. The failed LP refinements are preserved in the solve reports rather than counted as successful LP solves.

Each 52-period reference passed balance reconciliation and Equation (1); replay matched periods, result and world. Descriptions and worlds matched the corresponding no-decision and LLM cases; C1/C2 share a reference pair. Solver gaps concern planning programs, not engine-level global optimality.

The no-decision profit is −709331.2306 on each world; the archived myopic outcome is −432814.3424318 on each. For each baseline, all 52 decision submissions, execution outcomes, received deliveries and financial ledgers coincide between worlds. This explains equal profits despite differences in events and observations, without implying that the scenarios are interchangeable for other controllers. Equation (S1), defined in Section S4.2, gives $s=0.697555$ for B3 and 0.872800 for rolling on steady-market; on the outage-with-notice world, $s=0.570027$ for C1, 0.437927 for C2 and 0.554937 for rolling. Both denominators are positive. The outage-with-notice scores use a time-bounded heuristic anchor: two full-foresight planning steps reached the solve limit, as detailed above. The score is an affine scaling of profit within a world, preserving its ordering; its level depends on the anchors and is not an optimality percentage or an estimate of information value.

The no-decision loss comprises lost-sale penalties of 376982.56, late-delivery penalties of 274633.248, interest of 60586.1173 and storage costs of 24563.5156, offset by salvage of 27434.2103. This negative baseline makes positive scores possible even when profit is negative, as in C2. For fixed full-foresight profit, a more negative baseline reduces the score difference corresponding to a fixed profit difference. Direct matched-world differences against rolling are therefore the primary financial comparison in Table 3. They equal −316898.86142532 for B3, +24167.96939806 for C1 and −187403.31677434 for C2. These differences use model monetary units; they are not net benefits after model-inference costs.

Table 3 divides units shipped on time by all units ordered. B3's on-time/shipped rate is 27.70%, versus 33.12% for C1 and 35.96% for C2; these conditional rates have different denominators. None alone establishes which controller creates more business value. The rolling and LLM controllers also differ in planning procedure and computational budget. This small integration demonstration does not estimate their expected performance over worlds or model responses.

The references/ directory contains trajectories, baseline context, solver reports, audits, replay verdicts and the original protocol. Historical wrappers are not an entry point for the revised archive. Use the repository input audit and saved-action replay commands; the four 52-period references were not re-optimised during migration.

### Re-executing archived actions

The repository saved-action utility uses a temporary database and blocked external sockets, importing no model client or optimiser. It checks descriptions, submits stored decisions and compares every period, final result and complete world. The run-specific world endpoint returns the complete world only after completion.

On 13 November 2026, all eight cases passed from an extracted package with exact JSON-normalised periods, result, world and period counts: A3 (3 periods) and B3/C1/C2/B-rolling/B-reference/C1-rolling/C1-reference (52 each), totalling 367 periods. A control increasing archived

A3 profit by one was rejected. The 573 source files matched release hashes. This author-run check used the existing Python 3.14.10/macOS arm64 environment, not a fresh installation. It executed stored decisions without model calls or optimisation solves.

This validates the stored-decision replay path, not fresh controller effectiveness or stochastic sampling. Companion-utility corrections and validation reports are retained in historical/replay/validation/; the simulator was unchanged.

After installation and extraction as in Section S1.1, execute saved decisions from the repository root into a new output directory. This uses the repository implementation on the archived records:

```
uv run python -m scripts.reproduction.replay_saved_actions --case all --package-root "$PACKAGE"
--output data/replay-check
```

```
Individual cases can be selected with --case A3, B3, C1, C2, B-rolling, B-reference, C1-rolling
or C1-reference.
```

```
The output directory must not already exist.
```

README.txt gives current setup and individual-case commands. Request/response records are in integration/ and solver reports in references/. The historical author-run and Linux checks below are distinct from release CI and the 19 November repository validation.

At our request, an AI assistant (Claude, Anthropic) ran the eight saved-action replays on Ubuntu 24.04.5 x86_64 in a newly created virtual environment (uv sync --locked). Period records, final results and realised worlds matched exactly. Re-execution of 34 notice-matrix and 12 storage-sensitivity cases matched the same fields; differences were confined to timings and solver diagnostics, including about $2 \times 10^{-13}$ in one objective value and bound. This historical check concerns the earlier complete archive, not the revised data-only distribution or current repository client. Scope and commands remain in historical/linux-check/linux-check-report.md.

### S6. Evidence boundaries

Implementation claims and numerical evidence refer to the identified source snapshot. The related-software comparison is based on primary sources; classical process-management references support the modelling rationale.

The paper has no human-participant experiment, calibrated industrial factory or demonstrated adoption case. Prospective applications are not completed studies. Section S4.1 proposes a study protocol; it does not describe a campaign supplied with the repository. Sections S5.4–S5.5 supply author-generated integration and reference records. They describe observed behaviour without estimating comparative LLM-MAS effectiveness, information value or scenario difficulty.

### S7. Execution provenance

This section retains the staged verification and client-development history separately from the reproduction instructions. The archived reports preserve all attempts; the final client series is reported in full.

#### S7.1 Staged local verification

Local verification on 8–9 November 2026 cumulatively covered 3430 distinct non-stress items on macOS arm64/Python 3.14.10. Thirteen initial failures involved local permissions (ten) or an outdated uv (three). Twelve passed after environment corrections and one after a fresh Python installation directory, without source changes. This was staged coverage, not one uninterrupted run or hosted CI; stress tests were excluded. Section S1.2 identifies commands and reports containing the complete selections, failure causes, environments and timings.

### S7.2 Client development and run selection

The initial small-factory attempt stopped before closing any period because API quota was unavailable. The original local adapter completed A in 3 periods, whereas B, C1 and C2 stopped after 11, 3 and 7 closed periods on infeasible final plans. The revision-enabled v2 client completed A2, then stopped B2 after 18 periods on a procurement-call timeout in period 19; C runs were not started in that series. The timeout-handling v3 client completed A3 and B3. A Decimal serialisation error then stopped console reporting after B3's complete evidence had already been saved. A reporting-only correction resumed C1 and C2 without repeating A3 or B3 and without changing their decision logic. Both C runs completed. Earlier results and interruptions are retained in the accompanying records; the final series is reported in full rather than choosing its most profitable runs. This development history does not estimate an unchanged client's completion probability.

### S7.3 Reference preflight and companion-script checks

Two reference preflights stopped before period 1 on Decimal-versus-JSON-string comparisons; only the audit wrapper changed. Both records remain in references/.

The 14 November checks of run_allocation.py and make_figures_1_4_6.py both exited successfully in the existing macOS environment: seven allocation runs matched archived results, periods and worlds excluding solver timings, and the figures were rendered from archived inputs. Scope and reports are in historical/entrypoint-checks.json.

### References

Reference numbers follow the main article; references [1–27] are reproduced here for convenience.